\input psfig.sty 

\hsize=6truein
\overfullrule=0pt

\font\title=cmr10 scaled\magstep4
\font\subtitle=cmr10 scaled\magstep2
\font\author =cmr10 scaled\magstep1

\font\eaddr=cmtt10
\font\ninerm=cmr8
\baselineskip=16pt plus 1pt minus 1pt
\def\QED{\vrule height8pt width4pt depth0pt}

\def\proclaim #1: #2\par{\medbreak
	      \noindent{\bf#1:\enspace}{\sl#2}\par\medbreak}
\pageno=1	      

\hfill July 22, 2009\break

\vskip 1truein
\centerline{\title Dense packing on uniform lattices}
\vskip .7truein
\centerline{{\author Kari Eloranta}\footnote{$^*$}
{Research partially supported by The Finnish Academy of Science and Letters}}
\vskip .1truein
\centerline{\author Institute of Mathematics}
\centerline{\author Helsinki University of Technology}
\centerline{\author FIN-02015 HUT, Finland}
\vskip .1truein
\centerline{\eaddr kve@math.hut.fi}

\vskip .3truein

\centerline{\subtitle Abstract}
\vskip .3truein
\centerline{\vbox{\hsize 4.5in \noindent \ninerm \strut We study the Hard Core Model on the graphs ${\rm {\bf \scriptstyle G}}$ obtained from Archimedean tilings i.e. configurations in $\scriptstyle \{0,1\}^{{\rm {\bf G}}}$ with the nearest neighbor 1's forbidden. Our particular aim in choosing these graphs is to obtain insight to the geometry of the densest packings in a uniform discrete set-up. We establish density bounds, optimal configurations reaching them in all cases, and introduce a probabilistic cellular automaton that generates the legal configurations. Its rule involves a parameter which can be naturally characterized as packing pressure. It can have a critical value but from packing point of view just as interesting are the noncritical cases. These phenomena are related to the exponential size of the set of densest packings and more specifically whether these packings are maximally symmetric, simple laminated or essentially random packings. }}

\vskip .4truein
\centerline{\vbox{\hsize 4.5in \noindent Keywords: hard core model, independent sets, golden mean subshift,\hfill\break Archimedean tiling, densest packing, sphere packing \hfill\break
AMS Classification: 05B40, 52C15, 58F08, 82C26\hfill\break
Running head: Dense packing on uniform lattices}}

\vfill
\eject

\vskip .4truein
\noindent {\subtitle 0. Introduction}
\vskip .3truein

\noindent The original Hard Square Model has since its introduction in Statistical Physics surfaced in various other fields as well. Apart from modelling the interaction with infinitely strong but zero range potential (particles with no interaction except on contact, when they are undeformable), its exclusion rule of forbidding two 1's in neighboring graph vertices is natural in e.g. communication networks and in symbolic dynamics. In these contexts the model is often called independent sets or the golden mean subshift.

In this paper we concentrate on the packing aspects of the model, more specifically on the dense packing regime of its two-dimensional version. In short, we are interested in the qualitative properties of the densest packings and their formation in a discrete geometric set-up based on uniform tilings.

First we characterize the geometric and density properties of the optimal packings of 1's. After this we introduce a probabilistic cellular automaton that is parametrized by packing pressure. There may or may not be a dynamic phase transition as this parameter approaches its maximum. Through this formulation we will be able to see the connection between criticality and packing type. In the first case there is a phase transition and the densest packings form a finite set of laminated packings. The mechanism resulting in the critical behavior is in all our cases the same, a voter rule embedded in the hard core rule. The second, non-critical, case corresponds to the set of densest packings being exponentially large. This is due to the existence of a local move on the densest packings. In this class the optimal configurations are generically random packings. Finally there seems to be a qualitatively different \lq\lq borderline case\rq\rq\ corresponding to existence of a non-local move resulting in random laminated packings. To further characterize this case as well as some related phenomena we also investigate a few cases beyond uniform lattices.

This work is partly motivated by recent advances in packing in continuous space set-ups (see e.g. [CE], [CG-SS]). The overall picture from our 14 lattices compares interestingly with these results. Of particular interest is the fact that in ${\rm {\bf R}}^n,\ n\ge 1$ but \lq\lq small\rq\rq, the densest packings are believed to be lattice or random lattice packings (also in some exceptional dimensions like 24). On the other hand for high $n$ as well as in other \lq\lq big spaces\rq\rq\ like the hyperbolic space some type of random packings are expected to prevail. In our discrete set-up the random packings surface already in planar uniform graphs.

\vskip .4truein
\noindent {\subtitle 1.1. Set-up: the Rule and the Archimedean graphs}
\vskip .3truein

\noindent Let ${\rm {\bf G}}$ be a graph. In all but one of the cases considered in this paper it will be planar. We measure distance, $d$, on ${\rm {\bf G}}$ by hop count i.e. by computing the minimal number of edges that need to be traversed to move from vertex $x$ to $y.$ The (punctured) neighborhood of nearest neighbors of a vertex $x$ is $N_x=\{y\in {\rm {\bf G}}\ |\ d(x,y)=1 \}.$ $|N_x|$ is the vertex {\bf degree} of $x$ in ${\rm {\bf G}}$. 

Our configurations, which form a subset of $\{0,1\}^{\rm {\bf G}}$, are defined by the local 

\proclaim Hard Core Rule: If there is a $1$ at the vertex $x$ then on $N_x$ the configuration must be all-$0.$ \par

\noindent Denote the set of such legal configurations by $X_{\rm {\bf G}}^{hc}.$ 

\vskip .3truein
\noindent Most earlier studies on the hard core model have concentrated on the square lattice or more generally on ${\rm {\bf Z}}^d.$ Additional results exist on the triangular lattice, on trees and a few other set ups, see e.g. [B], [GK], [BW]. While the model on a graph is the most general one, it is not the natural set-up when one is interested in the packing aspect. Graph as a purely topological object needs to be augmented with a metric structure so that the nearest neighbors are in some sense also geometric neighbors.

${\rm {\bf Z}}^2$, ${\rm {\bf T}}$ (triangular lattice) and ${\rm {\bf H}}$ (honeycomb lattice) are the vertex and edge sets of the three {\bf regular tilings} of the plane. Their most immediate generalizations arise from the {\bf uniform} or {\bf Archimedean tilings}. While the regular tilings are constructed by tiling the plane with a single type of regular convex polygon, for uniform tilings a mixture of the different types of these polygons is allowed. Turns out that there are 11 such tilings of the plane. They share the common property that at each vertex the tiling looks up to rotation identical and the nearest neighbors (one edge away) are all at at the same Euclidean distance away which we take to be the unit. The code names of the tilings are obtained by circling a vertex and recording the $n$-gons along the way. Hence ${\rm {\bf Z}}^2$ is $4^4$, the triangular lattice is $3^6$ etc. For more on properties of the these tilings see [GS]. From now on we shall view the vertex and edge sets of the Archimedean tilings as geometric graphs, the {\bf Archimedean graphs/lattices} (most of them are not mathematical lattices but this terminology is in  use nevertheless). Figure 1. illustrates their set.

\vskip .4truein
\noindent {\subtitle 1.2. Densest packings}
\vskip .3truein

\noindent {\subtitle 1.2.1. Maximum density}

\vskip .3truein
\noindent Let us first investigate the density and geometric properties of the densest packings on the uniform lattices when the Hard Core Rule is imposed.

\vskip .2truein
\noindent The {\bf density of} $1$'s in a configuration $c$ is

$$\rho=\lim_{n\rightarrow\infty}{1\over |D_n|}\sum_{x\in D_n} c(x)$$

\noindent where the limit must exits and agree for all sequences $\{D_n\}$ retaining their two dimensional shape as $n\rightarrow\infty$ ($D_n$ is a domain of size $n$ lattice points on ${\rm {\bf G}}$). This number, when existing, is always between $0$ and $1$.

For some of our graphs it is immediate which sets support the densest packings (e.g. for ${\rm {\bf Z}}^2$ the checkerboard), for others it is less obvious. We first argue the optimal densities and present samples of corresponding configurations.

\centerline{\hbox{
 \psfig{figure=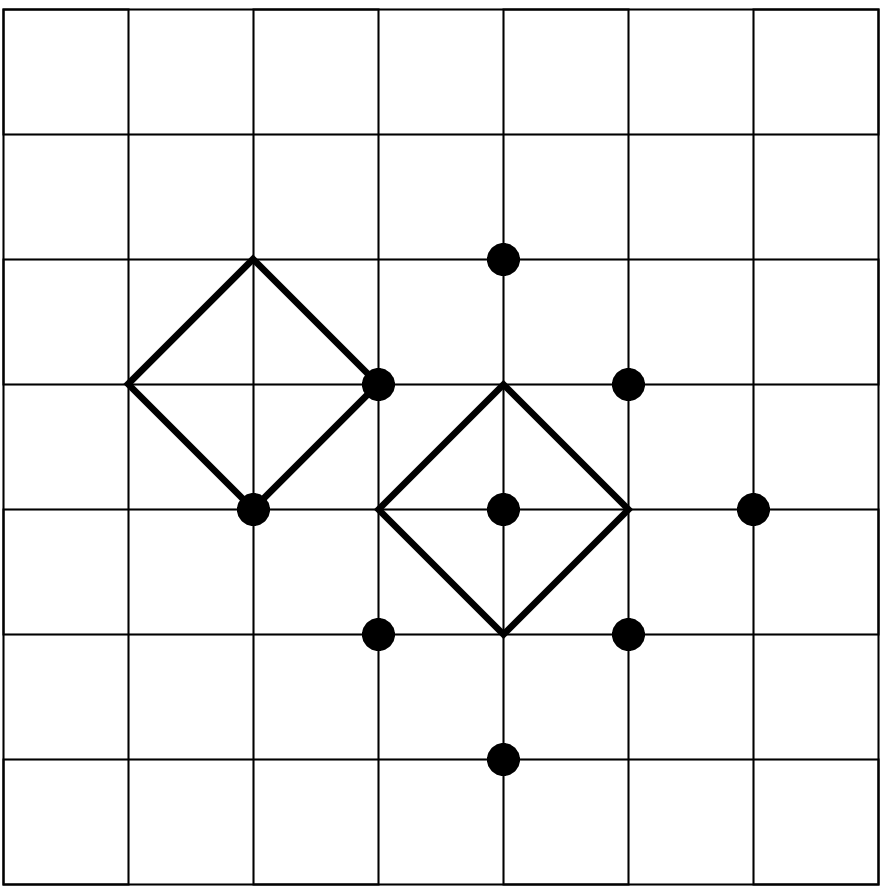,height=1.6in}
 \hskip .5truein
 \psfig{figure=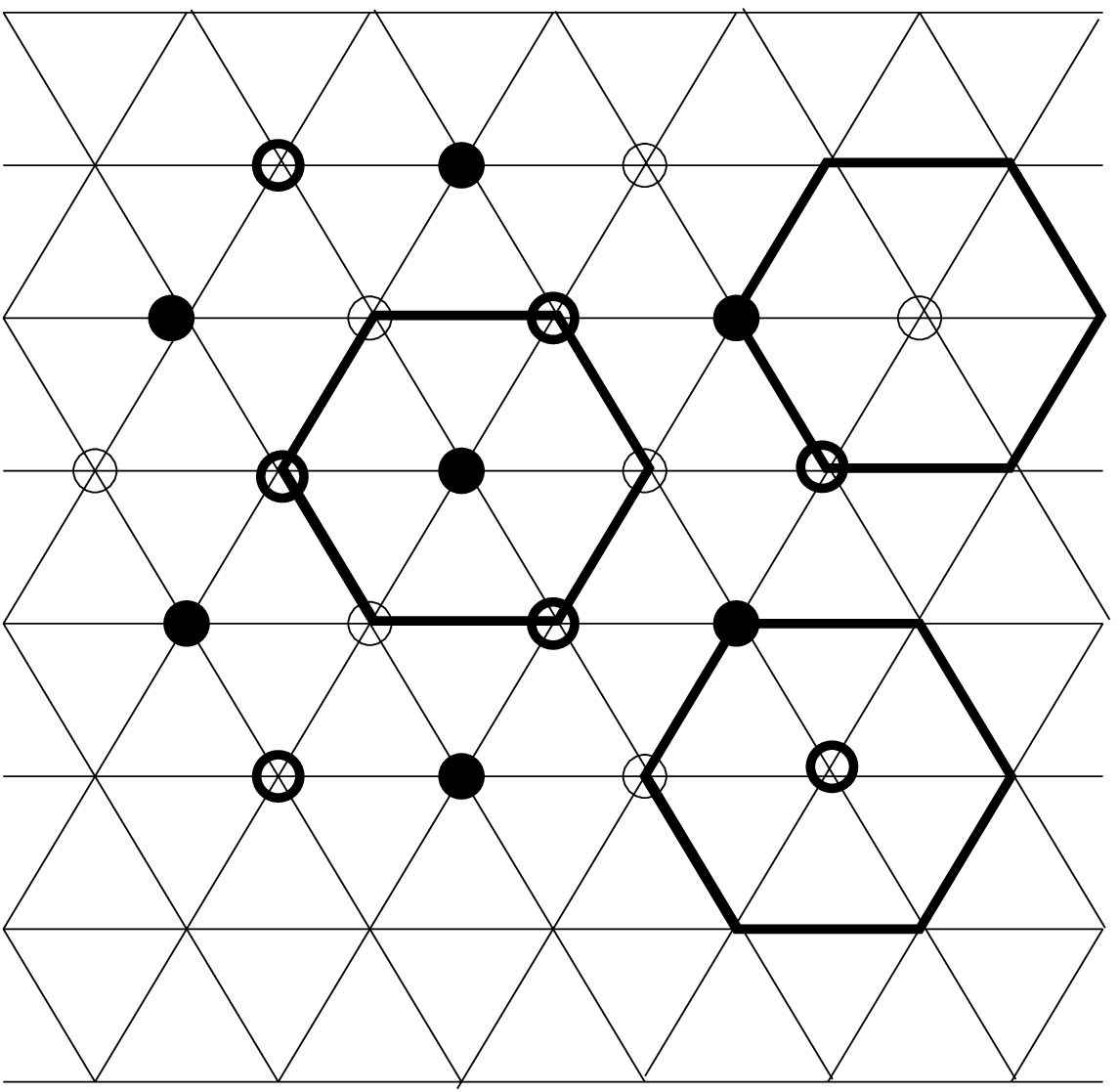,height=1.6in}
 \hskip .5truein
 \psfig{figure=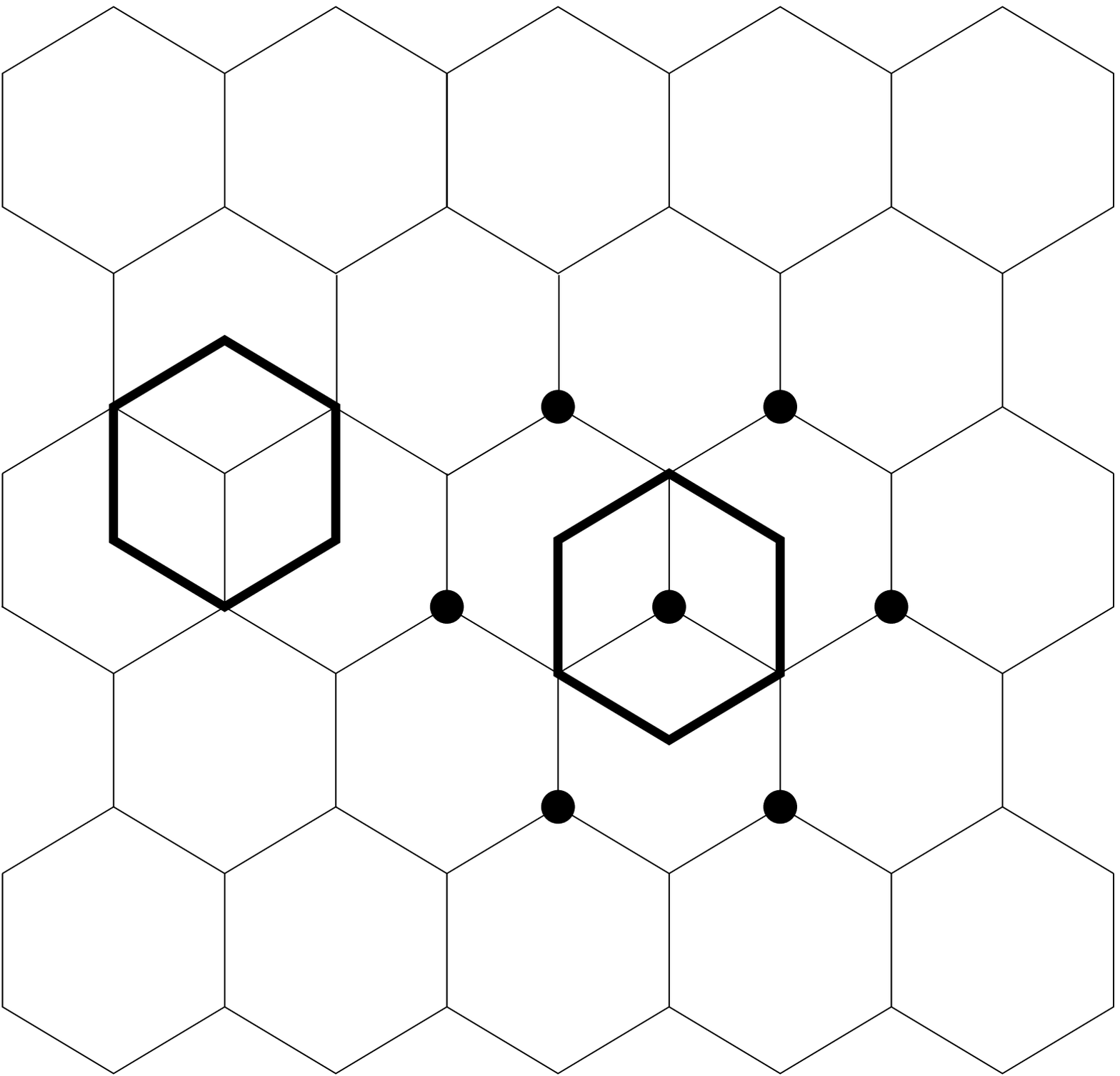,height=1.6in}
}}
\vskip .3truein
\centerline{\hbox{
 \psfig{figure=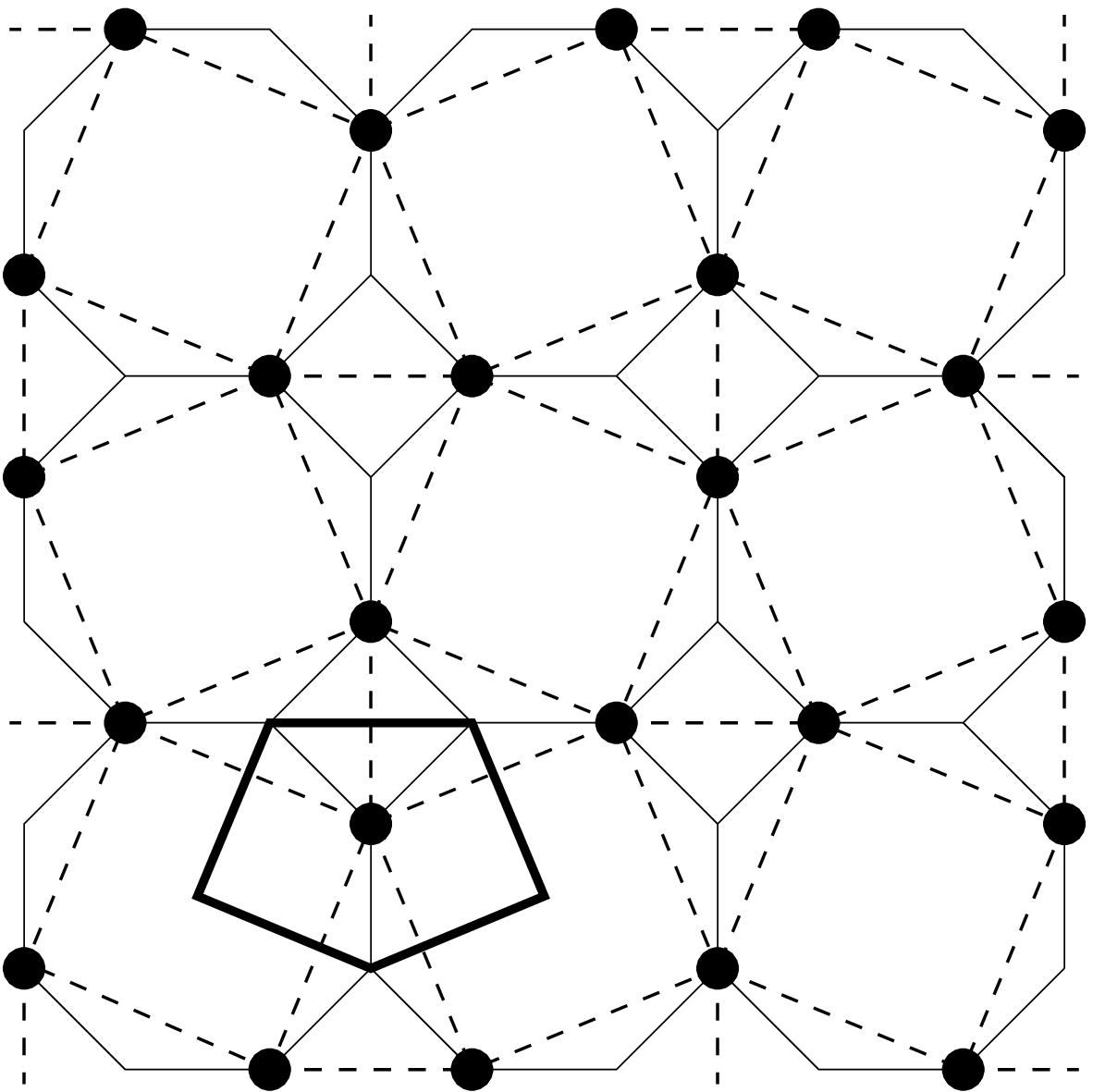,height=1.6in}
 \hskip .5truein
 \psfig{figure=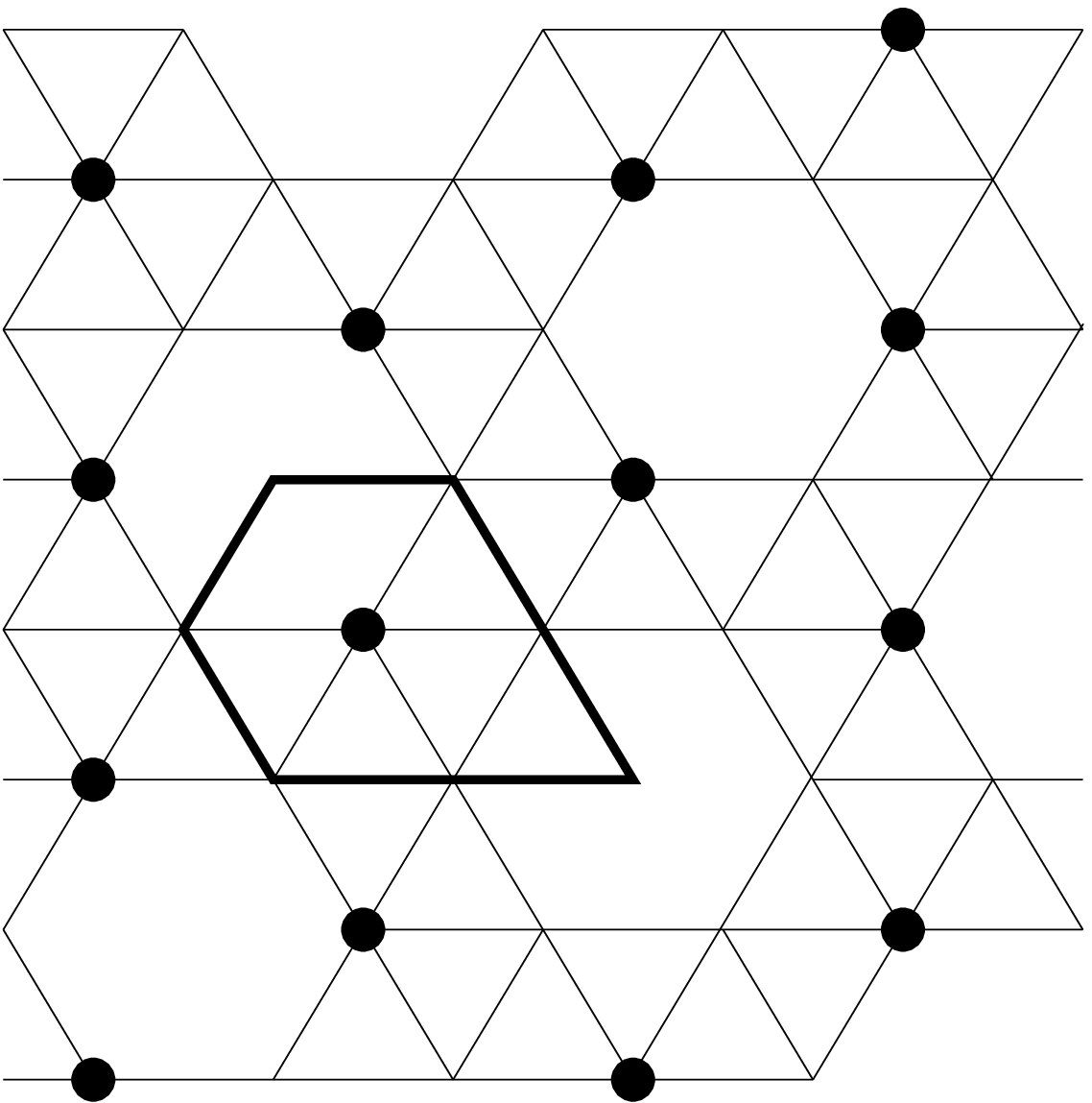,height=1.6in}
 \hskip .5truein
 \psfig{figure=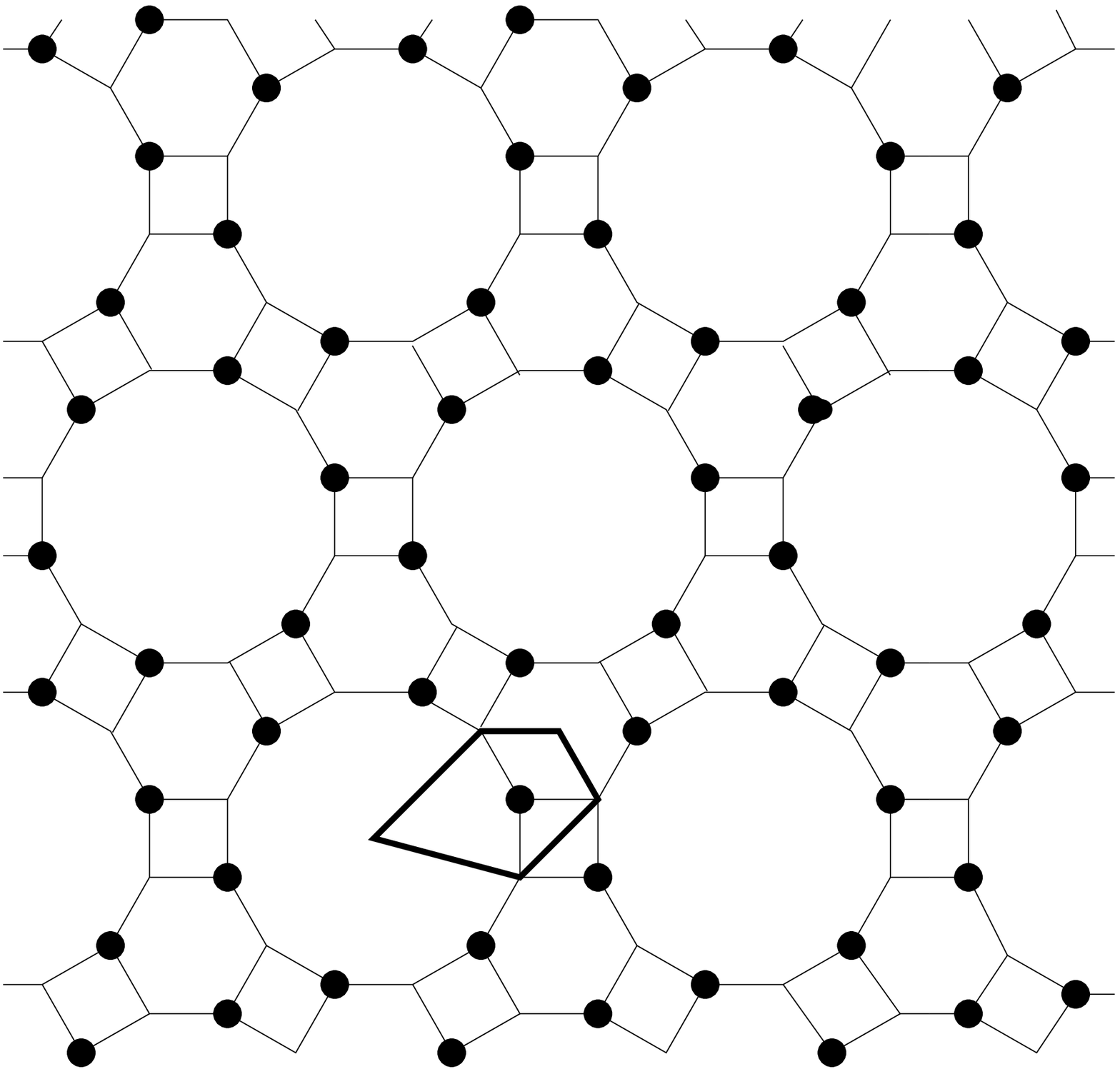,height=1.6in}
}}
\vskip .3truein
\centerline{\hbox{
 \psfig{figure=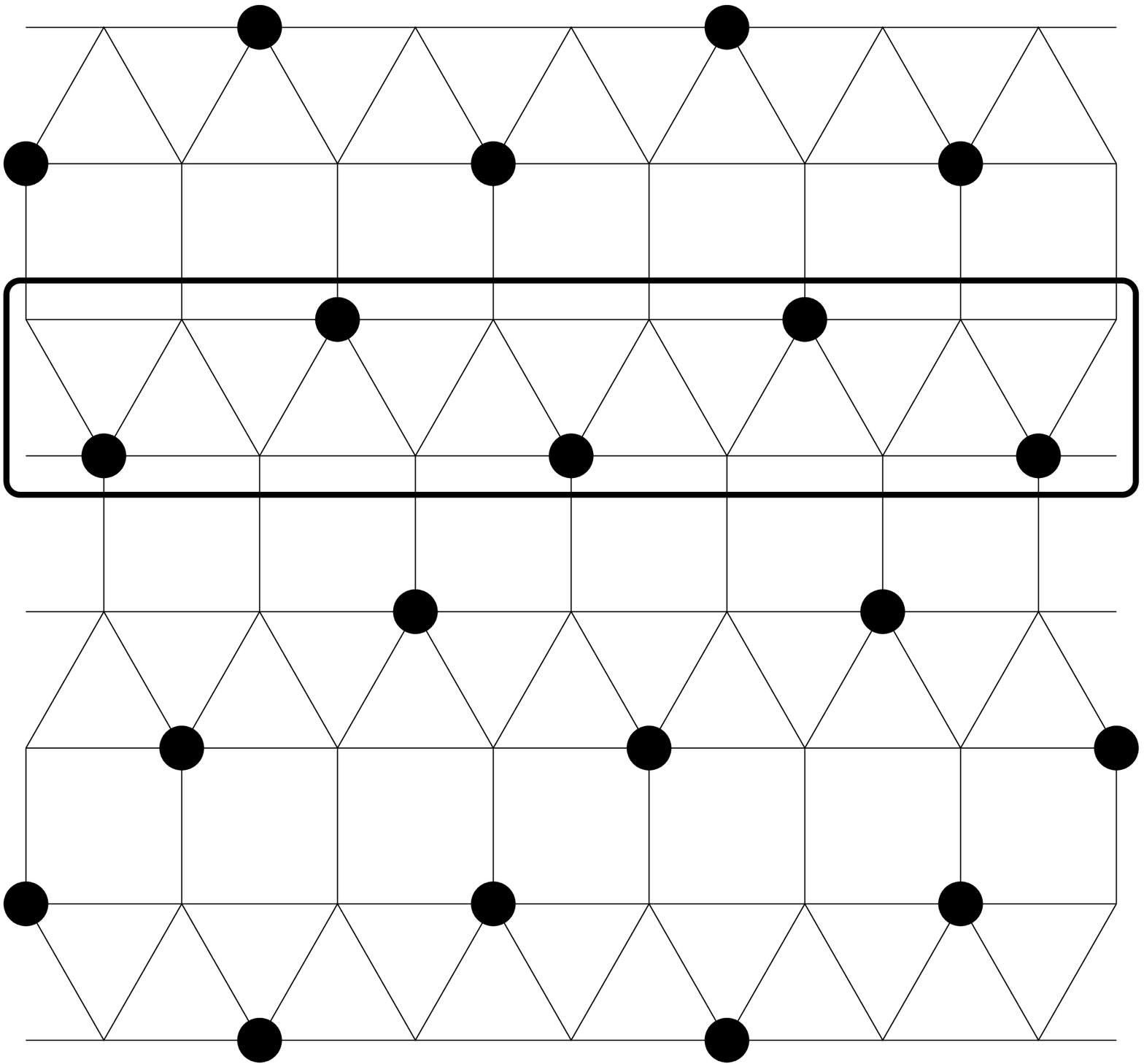,height=1.5in}
 \hskip .5truein
 \psfig{figure=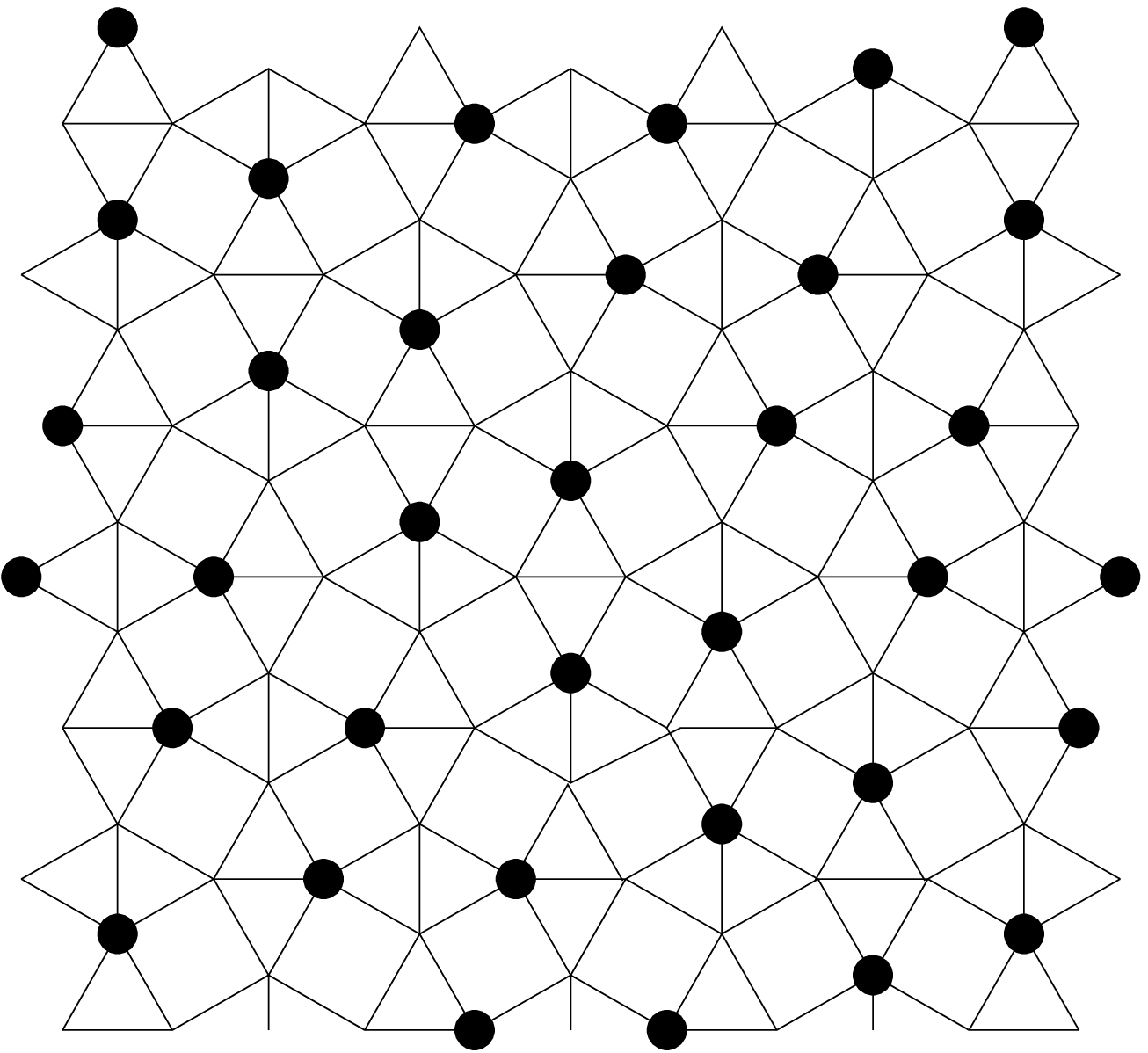,height=1.6in}
 \hskip .5truein
 \psfig{figure=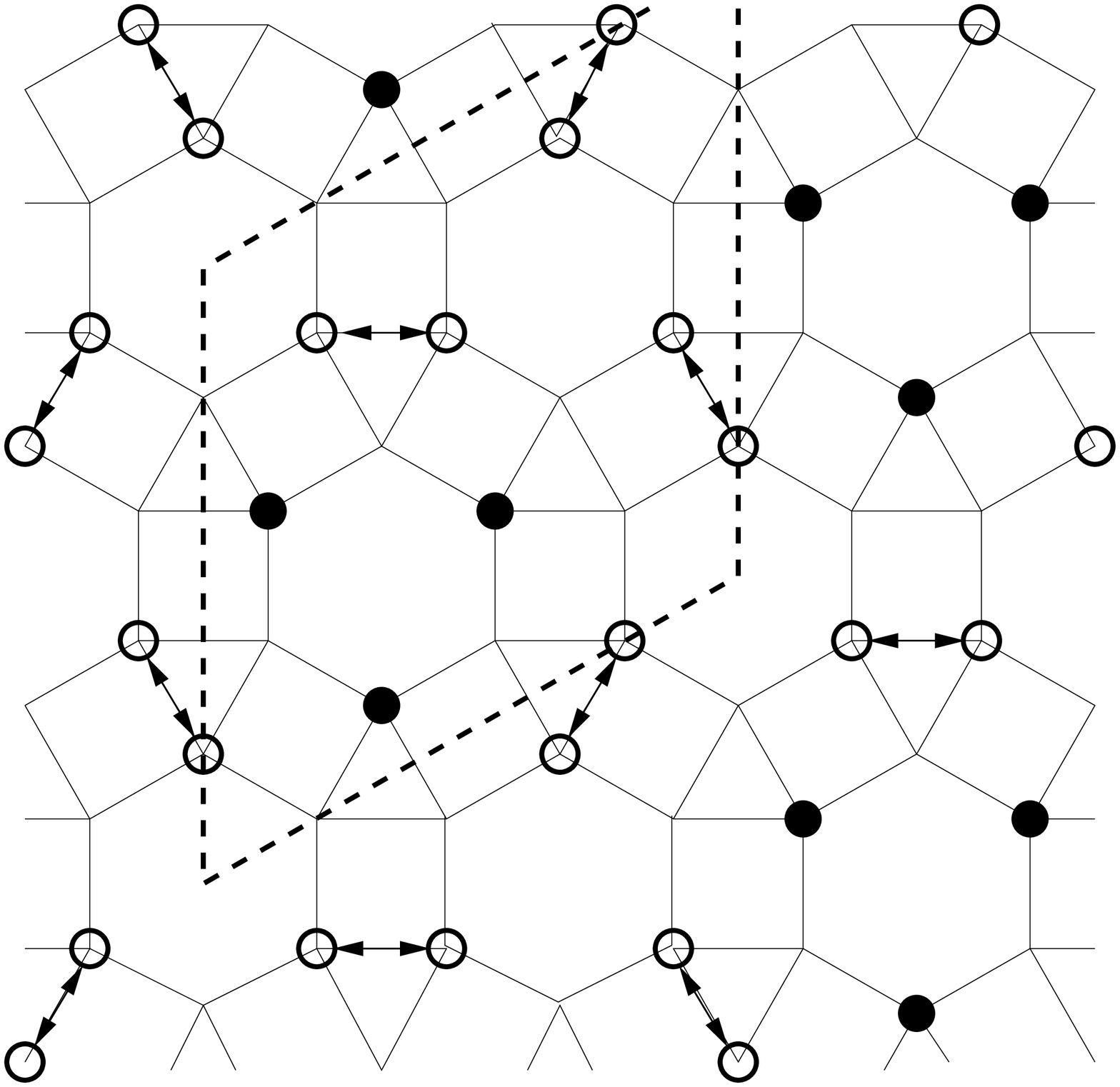,height=1.6in}
}}
\vskip .3truein
\centerline{\hbox{
 \psfig{figure=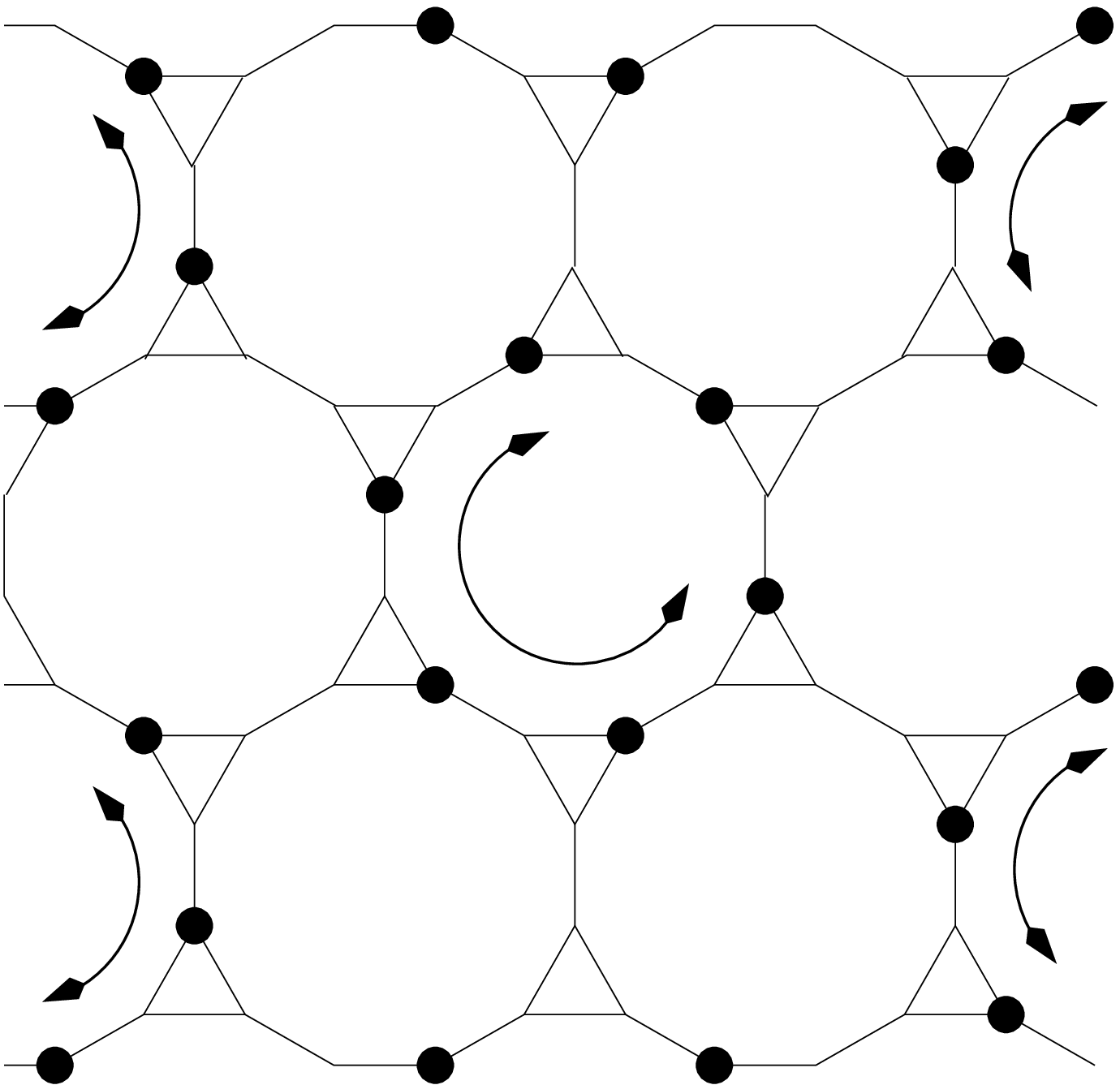,height=1.6in}
 \hskip .5truein
 \psfig{figure=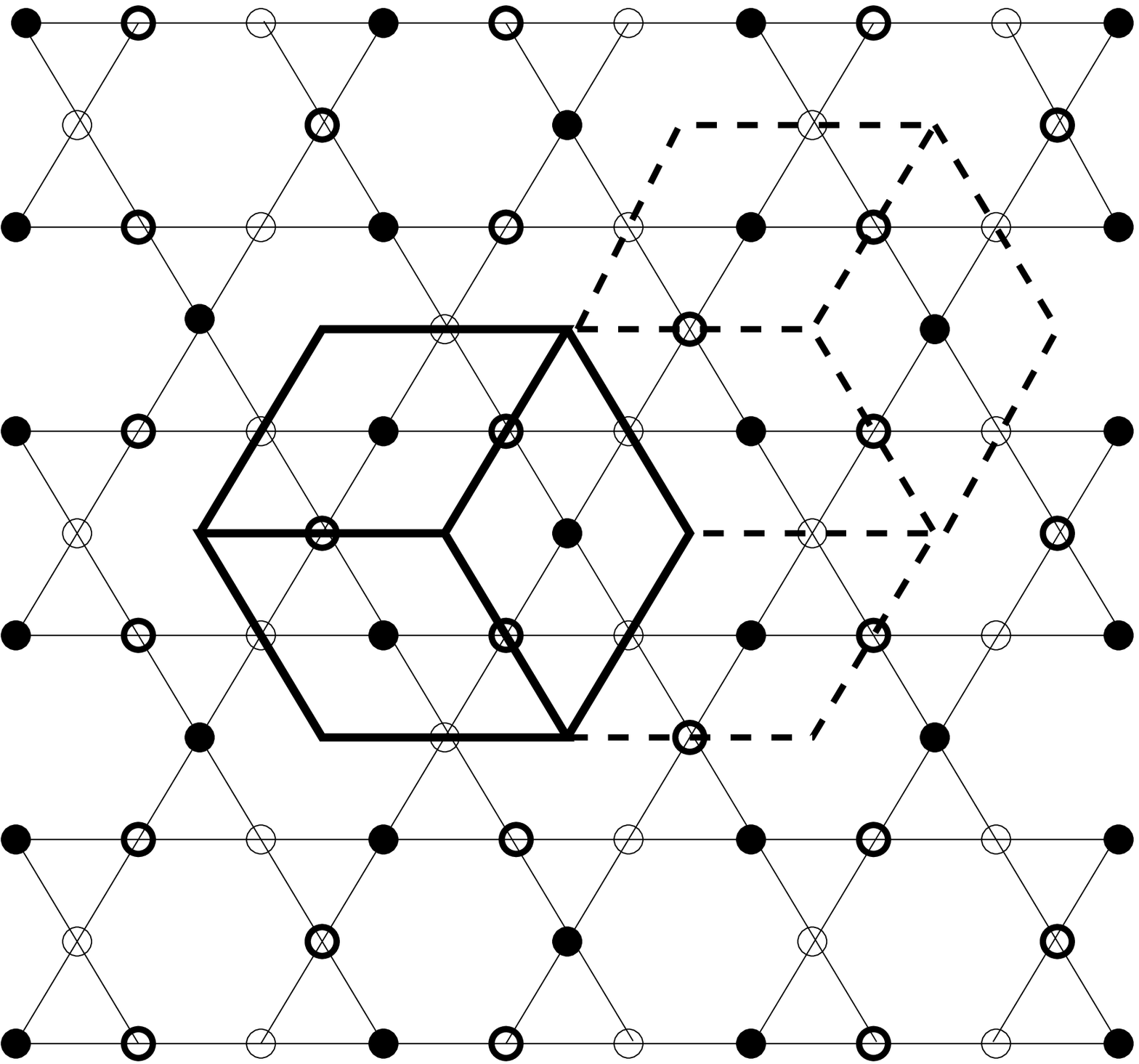,height=1.6in}
 \hskip .5truein
 \psfig{figure=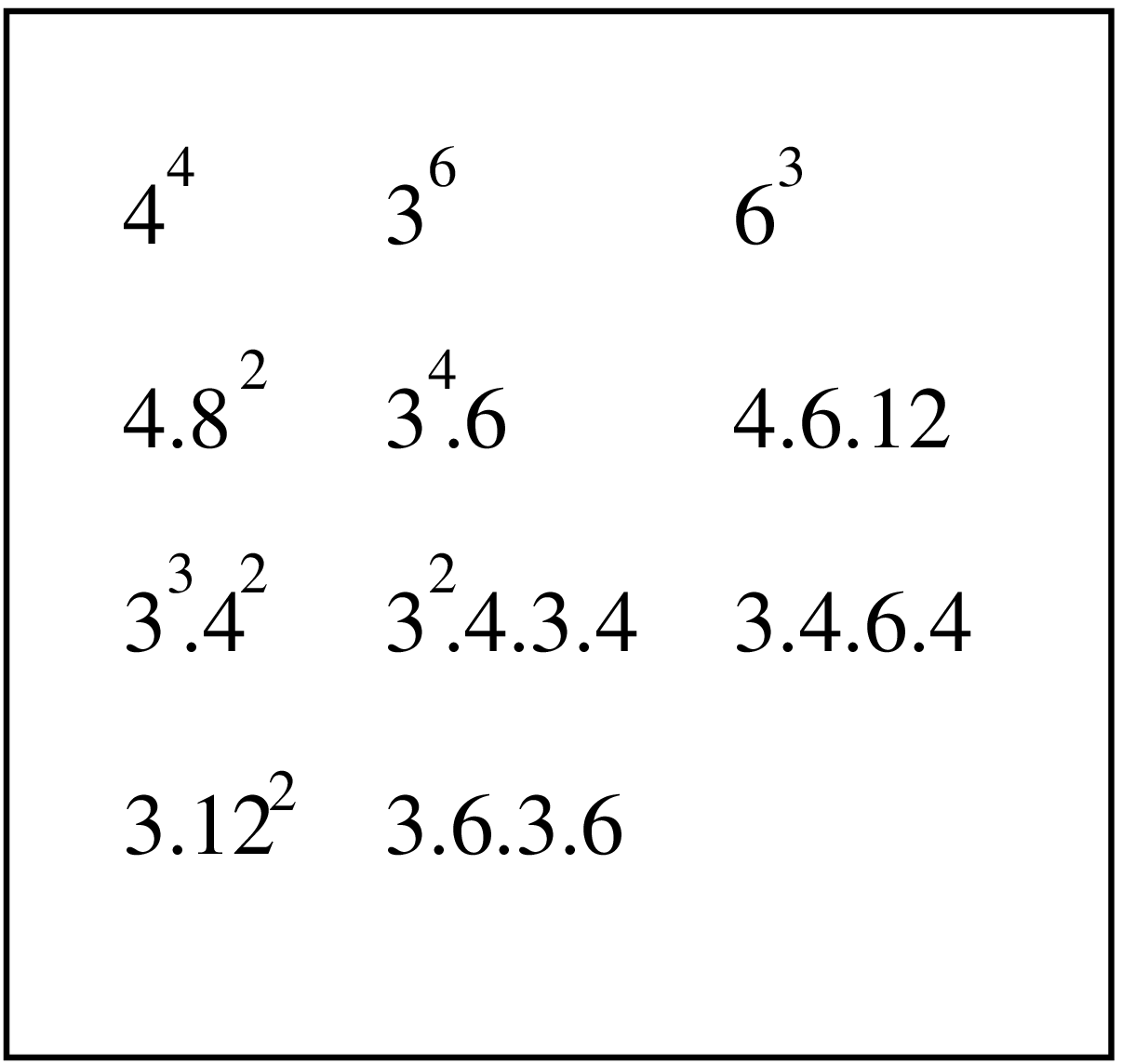,height=1.6in}
}}
\vskip .5truein
\centerline{{\bf Figure 1.} The Archimedean graphs.}
\centerline{The fine lines indicate the graph edges. For the details superimposed in bold see the text.}

\vfill
\eject

\noindent Consider a vertex $x\in G$ and the set of its second neighbors i.e. vertices two hops away, $N_x^{(2)}$. If $c(x)=1$ then by the rule on $N_x$ the configuration is all-$0$. On the set $N_x^{(2)}$ there may be both 0's and 1's. For the purposes of the upper bound argument we will ignore the configuration outside $\{x\}\cup N_x\cup N_x^{(2)}.$ For $y\in N_x$ let $n_y$ be the number of $y$'s nearest neighbor 1's inside $N_x^{(2)}$ and $n$ be its average:  
$$n={1\over {|N_x|}}\max_{\left\{{\rm legal\ conf.}\atop {{\rm on}\ N_x^{(2)}}\right\}}\sum_{y\in N_x} n_y \quad .\leqno{(1.1)}$$

\noindent Note that $n|N_x|$ always majors $\max|\{z\in N_x^{(2)}\ |\ c(z)=1\}|$ which is the discrete analog of the {\bf kissing number} of the sphere packing (see e.g. [CE]).

\proclaim Proposition 1.2.1.1: On an Archimedean graph of degree $d$ and second neighbor mean occupancy $n$ the global configuration density of $1$'s is bounded from above by
$${\overline \rho}={1\over {1 + {d\over n+1}}}\quad.\leqno{(1.2)}$$ \par

\noindent {\bf Proof:} For the $1$ at $x$ there are $d\ 0$'s neighbored by at most $dn\ 1$'s in $N_x^{(2)}.$ Hence the average number of $1$'s each $0$ is shared by is at most $n+1.$ Equivalently, on the average for each $1$ there are at least $d/{(n+1)}$ $0$'s. On an Archimedean graph $d$ is constant and since all neighborhoods are rotations of each other $n$ is constant, too. Hence the bound holds at every vertex on the graph and the upper bound follows. \hfill\QED  
    
\vskip .2 truein
\noindent {\bf Remarks: 1.} The result is actually more general than our set-up and also applies e.g. to homogeneous trees: $n=d-1$, hence ${\overline\rho=1/2}$ always. \hfill\break
\noindent {\bf 2.} Despite its simplicity the bound is often tight. For example for the square lattice we have $d=4$, $n_y\equiv 3$ hence $n=3$ and ${\overline\rho=1/2}.$ Indeed it is tight for the first five and last two Archimedean graphs as listed in Table I.

\vskip .2truein
\noindent For the remaining Archimedean cases the density bound is argued as follows.

\proclaim Proposition 1.2.1.2: For $(3^2.4.3.4),\ (3^4.6),\ (3^3.4^2)$ and $(3.4.6.4)$ the maximal density is $1/3.$ \par

\noindent {\bf Proof:} The infinite limit density bounds follow from the following arguments for configurations on a finite torus.

\noindent $(3.4.6.4):$ the graph is made of disjoint unit triangles (all unit edges) and since each can support at most one 1, the density is at most $1/3.$

\noindent $(3^2.4.3.4)$: each vertex is shared by three unit triangles, hence in the configuration the number of vertices is the same as the number of triangles. Since each triangle can carry at most one 1 the given bound follows.

\noindent $(3^4.6)$: each vertex is shared by four unit triangles, hence the number of triangles is $4/3$ of the number of sites. Therefore the number of 1's is bound by $1/4$ of the number of triangles which implies the bound $1/3.$

\noindent $(3^3.4^2)$: in each horizontal strip of triangles every 1 forces two 0's to its right implying the bound. \hfill\QED

\vskip .2truein
\noindent Examples of packings reaching the maximal density are shown in Figure 1 (1's are solid dots).

\vskip .4truein
\noindent {\subtitle 1.2.2. Qualitative features}
\vskip .3truein

\noindent Table I. lists some of the qualitative and quantitative packing properties of the Archimedean graphs (the first 11 lines) and three related set-ups that we analyze later. For ${\rm {\bf Z}}^2$ and ${\rm {\bf T}}$ existing results from the literature have been utilized but essentially all other data presented in Table I is established in this paper.

The first three lines correspond to the regular lattices. For example the first line indicates that the optimal packing on the square lattice $(4^4)$ is supported by one of two square lattices (with lattice spacing $\sqrt{2}$, tilted by $\pi/4$, hence diamond tile) which as sublattices each have density $1/2.$ As a packing it is simply what we would get by stacking the densest packings on one lower dimension, ${\rm {\bf Z}}$, therefore it is denoted by L for a {\bf laminated packing}.

The first seven cases are similar, call them {\bf rigid}: for each graph there is a (small) finite number of subgraphs all isometric to each other which support the densest packing and these arrangements of $1$'s cannot be deformed in any fashion. {\bf Topological equivalence} means that there is a homeomorphism between the subgraph and the stated Archimedean graph (in the terminology of [GS] the two graphs are of the same topological type). In all our cases the homeomorphism is actually close to the identity map so the actual subgraph \lq\lq looks like\rq\rq\ its preimage. See also Figure 1. for the geometry of the subgraphs involved ($1$'s are indicated by the solid dots).

 For $(3^3.4^2)$ the densest packing is an infinite family of {\bf random laminated/slide packings}. If every third entry in the 2-rows made of the triangles (framed in Fig. 1) carries a $1$ they can be stacked with off-sets $\pm 1/2.$ Hence in a square domain of $n\times n$ vertices there are approximately $e^{h^{(1)}n},\ h^{(1)}={1\over 2}\log{2}$ different packings all with $\rho=1/3$ (the superindex of $h$ refers to this being a 1-dimensional entropy).

The remaining three cases $(3.4.6.4)$, $(3.12^2)$ and $(3.6.3.6)$ (Kagom\'e lattice) are yet qualitatively different from those above. In all three the densest packings are essentially {\bf random packings}. They all have $\rho=1/3$ packings that allow {\bf local moves}: local exchanges of $0$'s and $1$'s between neighboring lattice sites (which preserve the legality of the configuration). Because of the locality of this action the number of packings in a domain of $n\times n$ vertices grows like $e^{h^{(2)}n^2}$, $h^{(2)}>0$ depending on the graph. For $(3.4.6.4)$ and $(3.12^2)$ see Fig. 1 for the details of these exchanges (the arrows).

\vskip .2truein
\noindent Subsequently we call maximally symmetric densest configurations {\bf optimal packings}. For laminated packings these are all there are, for the other two types they are rare specialties.

\vskip .2truein
\noindent Having identified a densest packing subgraph for an Archimedean graph and viewing it as a subset of ${\rm {\bf R}}^2$ we can define its Voronoi tessellation. Every $1$ is the center of a convex {\bf 1-tile} which is unique up to rotation. If the subgraph is in fact one of the Archimedean graphs this tile has the $0$'s of the set $N_x$ along its edges. An identical tile with the symbol $1$ at the center replaced by $0$ is a {\bf 0-tile}. In Table I we have identified the types of these tiles and some of them are illustrated in Figure 1 (in bold line). Archimedean graphs for which this construction yields the simplest tiles are $(4^4),\ (6^3),\ (3^6)$ and $(3.6.3.6).$ Note that considering the packing of these 1-tiles on ${\rm {\bf R}}^2$ instead of $1$'s on ${\rm {\bf G}}$ does not introduce an extra condition. Because of the Hard Core Rule there cannot be 1-tile overlaps. If a tile arrangement in plane is in fact a tiling i.e. a one-fold cover of the plane then it is necessarily the densest packing of $1$'s. By the arguments above the rigidity properties of the densest packings translate into the rigidity of these tilings.

\vskip .4truein
\noindent {\subtitle 2.1. Dynamical packing with a PCA}
\vskip .3truein

\noindent We now introduce a dynamics that enables us to generate increasingly dense packings in all our set-ups. This yields further insight into their nature and in particular will clarify the nature of the three qualitatively different packing classes identified in the previous section. 

Compatible with the Hard Core Rule the configurations can be changed locally according to the following local

\proclaim Update: If there is at least one $1$ on $N_x$ then the symbol update at $x$ is $0$, otherwise it is $1$ with probability $p$ independently of the updates outside $N_x\cup \{x\}$. Denote this random local map from $N_x$ to $\{0,1\}$ by $f_p.$\par

\noindent Instead of the probability $p$ a related quantity called the activity (fugacity, $z=p/(1-p)$) can be used. Sometimes it is taken to be sublattice dependent corresponding to sublattice preference (akin to an external field in the Ising model). We do not consider this variation; our rule is uniform over the lattice. Its a simple but important result that the update, when nontrivial, is an irreducible action on the configurations:

\proclaim Proposition: For $0<p<1$ the Update can transform any legal finite configuration to any other such configuration with a finite number of steps. \par

\noindent {\bf Proof:} Given any legal finite configuration, there is a positive probability that it will be transformed by the update in one step to the all-0-configuration. Since this operation is reversible for the given $p$ any two such configurations can thus be transformed to each other with positive probability in two steps. \hfill\QED

\vskip .2truein
\noindent An efficient way of generating all legal configurations from any given configuration can be given once the subgraphs carrying the optimal configurations have been identified (as in Table I):

\vskip .1truein
\item{\it 1.} {\it Take any initial configuration on $G$ (legal or not).}
\item{\it 2.} {\it Pick any optimal subgraph and apply the Hard Core Update at neighborhoods centered at its vertices to generate an updated configuration on the subgraph.}
\item{\it 3.} {\it Cycle through all the optimal subgraphs in turn using the rule in 2. Once this is done all vertices of the graph have been updated.}
\item{\it 4.} {\it Combine the subgraph configurations into a global configuration on $G$ (which will now be legal).}

\vskip .1truein
\noindent The global map $F_p:X_{\rm {\bf G}}^{hc}\rightarrow X_{\rm {\bf G}}^{hc}.$ defined by steps 2-4 is our Probabilistic Cellular Automaton (PCA). Iterating it will relax the initial configuration to a Gibbs state on $X_{\rm {\bf G}}^{hc}$ (a measure of maximal entropy on configurations compatible with the Hard Core Rule).

\vskip .1truein
\noindent {\bf Example:} On ${\rm {\bf T}}$ the three sublattices are shifts and tilts of $\sqrt{3}{\rm {\bf T}},$ call them dot, circle and ring sublattices (optimal sublattices, identified this way in Figure 1). If we take e.g. all-1 initial state on  ${\rm {\bf T}}$ (illegal) and choose first to update the dot sublattice, it will become all-0 independent of $p$ (since circle and ring are still all-1 and $N_x$ consists of them for each $x$ in dot sublattice). In step 3 dot and circle will in turn update the ring sublattice the same way to all-0 (due to 1's on circle). This is followed by circle being updated by dot and ring according to the probability $p.$ After the first full cycle through the three sublattices the subconfigurations combine to a legal configuration.

\vskip .1truein
\noindent Among our graphs either two or three sublattices are involved in the iteration except for ${\rm {\bf Z}}^2{\rm M},$ which cycles through four sublattices. 

\vskip .2truein
\noindent By the Proposition the action of the non-trivial PCA is irreducible on the set of Hard Core configurations. Hence in particular this algorithm can be used to generate all configurations from one sample (in a finite number of steps for a finite lattice).

It is natural to view the update probability $p$ as the {\bf packing pressure}: the PCA with higher $p$ is just simply squeezing more $1$'s into the available slots. In terms of the auxiliary tilings just introduced we fill in new $1$-tiles centered on the subgraph being updated if the rest of the $1$-tiles allow i.e. if no overlaps result.

\vskip .4truein
\noindent {\subtitle 2.2. Criticality and noncriticality in packing}
\vskip .3truein

\noindent The PCA above has been coded for our set of lattices and we will present here the results from the relaxation study. Apart from the square and triangular lattice cases which can be found in the literature the subsequent data is from our runs.

The aim here has not been great numerical accuracy but rather in verifying the critical behavior with a reasonable estimate for the critical parameter $p$. We have used the following sequential procedure. First we relaxed on a small (about $50\times 50$ sites) toral lattice to see if $p_c<1.$ If so we then obtained a better estimate for $p_c$ on a larger toral lattice (at least $200\times 200$). Finally we employed \lq\lq worst case\rq\rq\ initial conditions. In the subcritical regime we relaxed initial conditions filling completely one of the sublattices to see if the sublattice densities converged to a common value. In the supercritical case we used Bernoulli initial distribution with density of 1's approximately $\rho(p_c)$ and observed if the densities on the sublattices asymptotically diverged. This procedure was done with an increasing/decreasing sequence of $p$'s in the sub/supercritical case respectively to obtain a bracket for $p_c.$

\vskip .3truein
\noindent {\subtitle 2.2.1. Regular and other lattices with laminated packings}
\vskip .2truein 

\noindent The model on ${\rm {\bf Z}}^2$ is sometimes referred to by the name the Hard Square Gas. The squares are actually diamonds of side length $\sqrt{2}$ that we have identified above, in Table I and in Figure 2. The critical value for $p$, $p_c\approx 0.79$ was computed to high accuracy in [BET] and the Dobrushin-Shlosman criterion for phase uniqueness was applied in [RS] to provide a rigorous lower bound. Below the critical probability the model is ergodic, above the dynamics preserves two Gibbs measures assigning different probabilities on the two sublattices.

\vskip .3truein
\centerline{\hbox{
 \psfig{figure=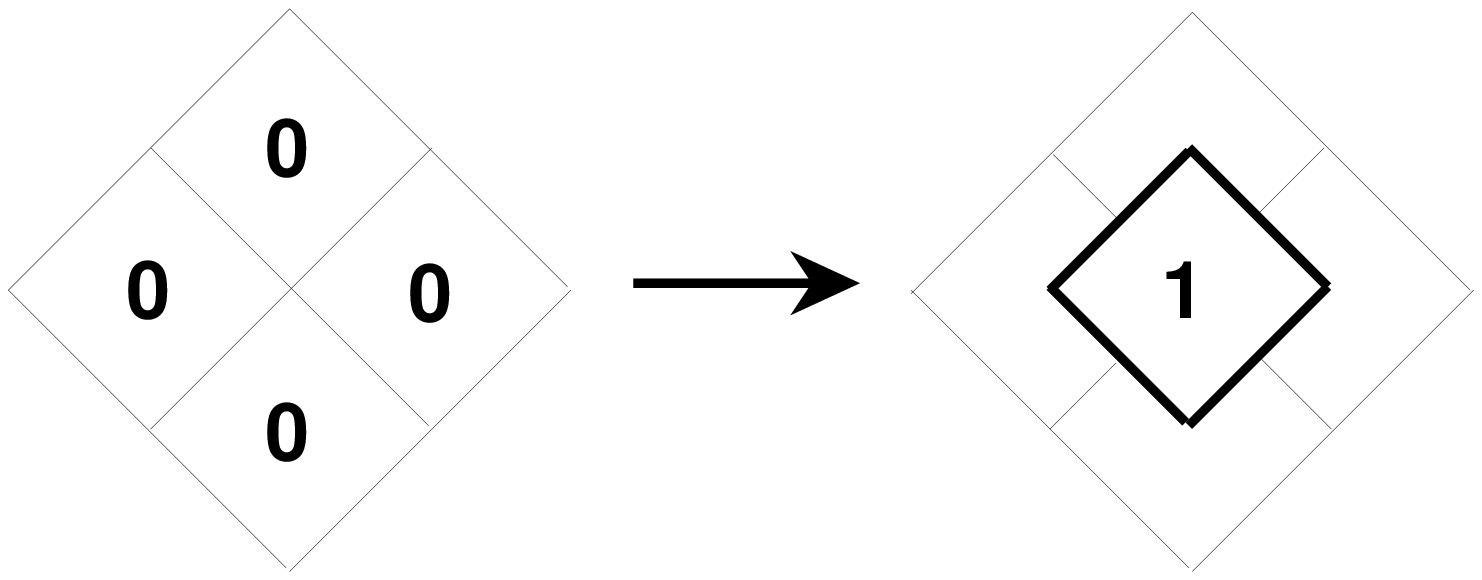,height=.8in}
 \psfig{figure=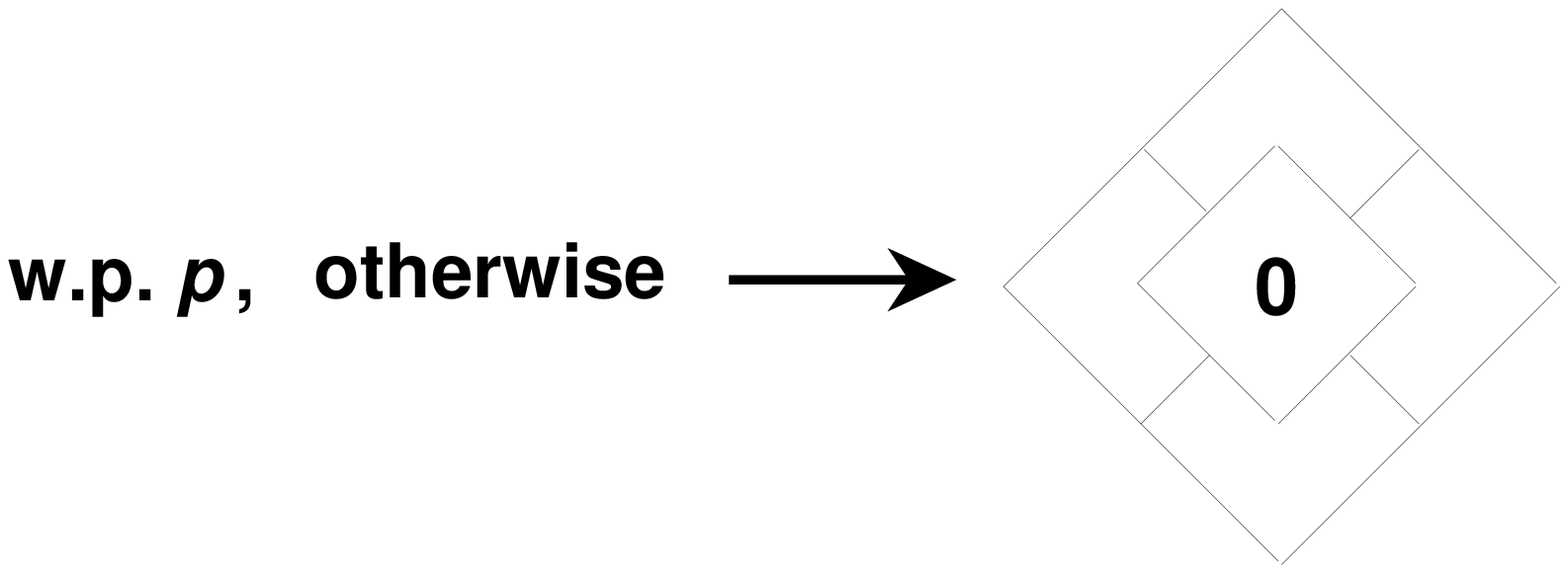,height=.8in}
}}
\vskip .1truein
\centerline{{\bf Figure 2.} The Probabilistic Cellular Automaton on ${\rm {\bf Z}}^2$ diamond tiles.}
\vskip .2truein

\noindent On the honeycomb lattice ${\rm {\bf H}}$ the dynamical behavior is similar to the square lattice case (for criticality proof see [R]). Here the geometry is as indicated in Figure 1: hexagonal $1$-tiles on two triangular lattices. In the supercritical case the densities of 1-tiles are unequal on the two sublattices whereas in the subcritical case they converge to a $\rho(p)$. Figure 3. shows a typical relaxation of these densities in the subcritical regime. The mirror-effect in the graphs is due to the fact that a high packing density ($\rho\approx 0.384$) has been attained and preserved in spite of the fluctuations around the equilibrium level.

\vskip .4truein
\centerline{\hbox{
 \psfig{figure=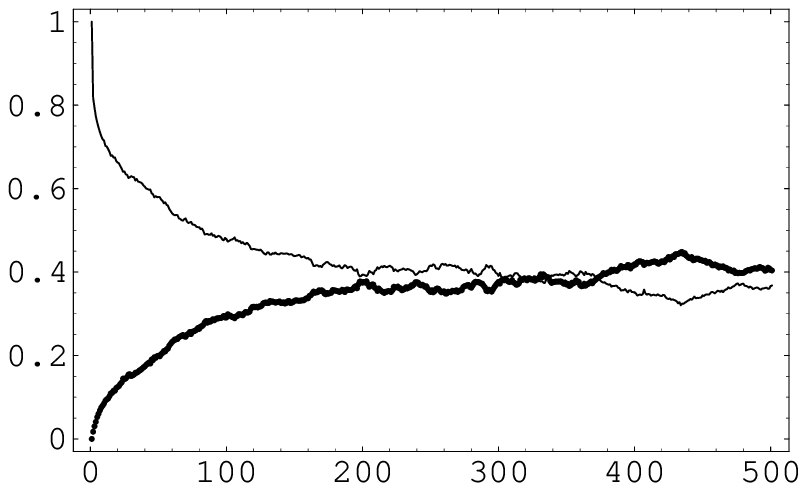,height=1.8in}
}}
\vskip .1truein
\centerline{{\bf Figure 3.} Relaxation on the honeycomb lattice ($p=0.859$, $200\times 200$ torus, 5.000 iterates.)}

\vskip .2truein
\noindent On the triangular lattice ${\rm {\bf T}}$ the model corresponds to the Hard Hexagon Model of [B]. It is known that beyond $p_c\approx 0.92$ there is a three way sublattice split into identical but thinner triangular lattices supporting the hexagons. Our runs indicate that the highest density of a packing that does not yet reveal the sublattice parity is approximately 0.26.

Our simulations on $200\times 200$ site (sometimes bigger) toral lattices established that on the lattices ($4.8^2$) and (4.6.12) the two sublattice alternatives carrying the densest packings emerge above critical values $p_c\approx 0.9$ and $0.92$ respectively. In the case of lattices $(3^2.4.3.4)$ and $(3^4.6)$ three way splits take place approximately at the critical probabilities $0.99$ and $0.97.$ The corresponding critical densities are listed in Table I.

\vskip .3truein
\noindent {\subtitle 2.2.2. Embedded voter rule}
\vskip .2truein

\noindent The critical behavior in the cases above seems to result from a common underlying structure. In short in these models one can identify a sublattice on which the square of the update is behaving essentially like a slightly asymmetric {\bf voter rule}.

Suppose that we have configurations from $\{0,1\}^{{\rm {\bf G}}}$ and a finite neighborhood $N.$ The classical majority voter rule has the following basic properties (see e.g. [L]).
\vskip .1truein
\item{$\bullet$} If $N$ has $n_s$ copies of $s$ then the update to this symbol takes place with a non-decreasing probability $p(n_s).$ If $n_s>|N|/2$ then $p(n_s)\ge n_s/|N|$ i.e. the update favors the majority. The update is symmetric in the symbols and if $|N|$ is even then in case of a tie in $N$ the symbol is updated without a bias i.e. w.p. $1/2.$

\item{$\bullet$} The updates in disjoint neighborhoods are independent. 

\vskip .1truein
\noindent Intuitively these mean that the interpolated graph of the function $p(n)$ is \lq\lq $\int$-shaped\rq\rq, symmetric with respect to $(|N|/2,1/2)$ and hits the corners $(0,0)$ and $(|N|,1).$

\vskip .2truein
\noindent Consider the map $f_p^2$ on the honeycomb lattice. It is a random map from a 7-tuple to the set $\{0,1\}.$ The 7-tuple is the dotted subset on the $\sqrt{3}{\rm {\bf T}}$ sublattice as illustrated in Figure 1. Similarly $f_p^2$ e.g. for ${\rm {\bf Z}}^2$ is supported by a $3\times 3$ diamond on a $\sqrt{2}{\rm {\bf Z}}^2$ sublattice.

Now take $p=1$ and plot the fraction of neighborhoods that yield update $1$ under $f_1^2$ as a function of the number of $1$'s in the neighborhood (the 7-tuple or $9$-tuple identified above). These have been plotted in Figure 4a and 4b, the leftmost graphs in each. The graphs clearly have the same qualitative character as the voter curves except that they favor the update $1$ even stronger resulting in slight asymmetry.

In the triangular lattice case the set-up is slightly more delicate. In Figure 1 the dots, circles and rings in a hexagonal array identify the $19$-tuple from which the update to the center dot is determined. In this case we consider the restriction $f_\bullet^2$ of $f_1^2$ to the dot sublattice. This map from a $7$-tuple is again a voter-type map as seen in Figure 4a, right. The other sublattices work identically.

\vskip .4truein
\centerline{\hbox{
 \psfig{figure=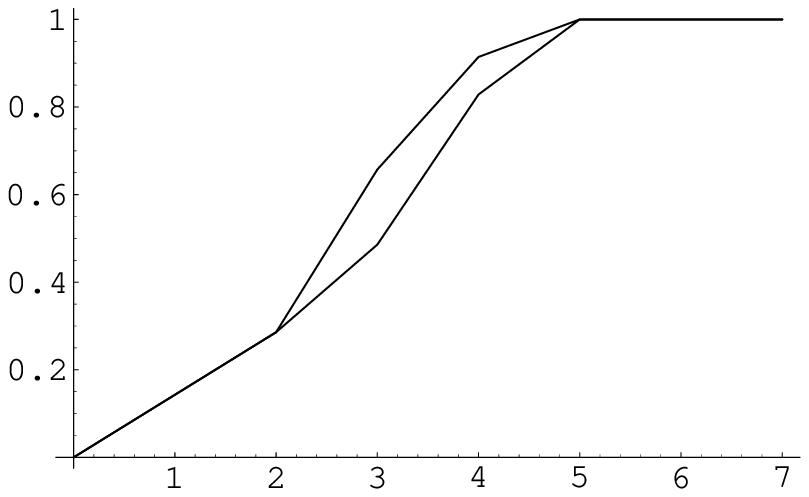,height=1.3in}
 \hskip .5truein
 \psfig{figure=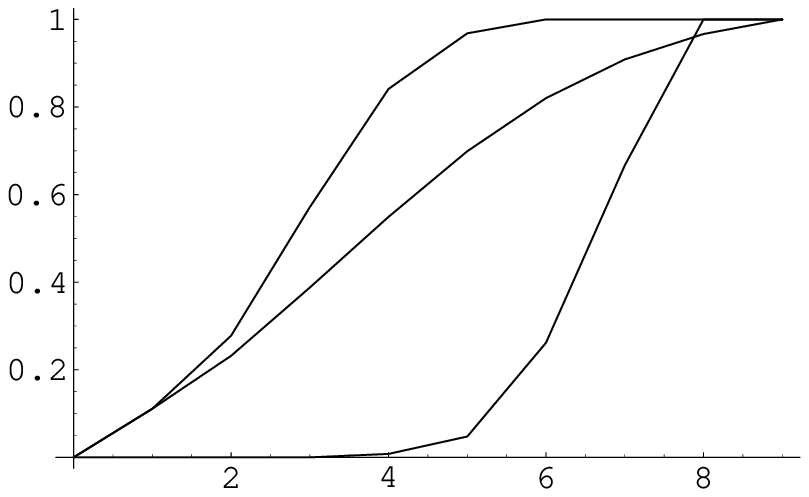,height=1.3in}
}}
\vskip .1truein
\centerline{{\bf Figure 4a, b.}  Voter curves for ${\rm {\bf H}}$, ${\rm {\bf T}}$, ${\rm {\bf Z}}^2$ and ${\rm {\bf Z}}^2{\rm M}$ (2) graphs (left to right)}
\vskip .2truein

\vskip .2truein
\noindent The key idea here is that the voter rule is known to be critical ([L]) i.e. depending on the steepness of the graph of $f^2$ around $|N|/2$ there is $p_c<1$ s.t. above it the dynamics becomes nonergodic. 

It is easy to see that for $f_p^2,\ p\approx 1,$ the update probability closely approximates the three curves above. Each one of them is for high $p$ a rule that favors the $1$'s to take over one of the sublattices thereby blocking the $1$'s from the other sublattices. Hence the criticality in these models seems to coincide with the criticality of the majority voter rule.

\vskip .4truein
\noindent {\subtitle 2.2.3. Slide packings}
\vskip .3truein

\noindent Among the Archimedean graphs our model on $(3^3.4^2)$ is exceptional. Table I. indicates that its high density characteristic property is the randomness in the lamination i.e. stacking of $3{\rm {\bf Z}}$ densest packings. We now investigate this phenomenon by going into a slightly richer lattice set-up.

\vskip .2truein
\noindent Let us reconsider the ${\rm {\bf Z}}^2$ lattice with the Moore neighborhood i.e. with neighborhood set extended to the nearest eight neighbors in the Euclidean distance on ${\rm {\bf R}}^2.$ The neighborhood graph is not planar anymore since the graph edges of the diagonals do not intersect.

In the hard core model on this graph the $1$-tiles are $2\times 2$ squares. There are four maximally symmetric tilings each supported by a copy of $2\ {\rm {\bf Z}}^2$. They are optimal packings but not the only ones. In such a packing we can pick any column/row with a $1$ in it and shift the entire column/row by $\pm 1$ lattice sites. This is exactly the kind of behavior that distinguishes the model on $(3^3.4^2),$ too.

But ${\rm {\bf Z}}^2{\rm M}$ is a bit more subtle. After shifting one column we can still slide in similar fashion any other column but none of the rows. This obviously generates an infinite number of configurations that are all densest possible. The one-dimensional entropy
$h_1=\lim_{n\rightarrow\infty}{1\over n} \ln{2^{n/2}}={1\over 2}\ln{2}$
is the exponential growth rate of the number of these configurations from infinite strips of width/height $n$ lattice sites.

In terms of packing the optimal ones are still all laminated packings but somewhat nontrivially so. A pair of sublattices that are exactly one unit lattice shift (vertical or horizontal) from each other can together support a perfect tiling whereas a pair of sublattices that are one horizontal and one vertical unit shift from each other can not. Because a random mixing of the two sublattices is allowed (as in the counting above) now essentially all of the optimal packings are disordered laminated packings. Some of their (non)ergodic properties were analyzed in [E].

\vskip .2truein
\noindent From the above we can expect that the equilibrium of the PCA evolutions when $p\uparrow 1$ is more complex than in the previous cases. This is indeed the case and in particular understanding the critical behavior becomes more challenging. There are an infinite number of ground states but in fact the only critical transition is between the two possible \lq\lq orientations\rq\rq\ of the ground state. This has caused significant difficulty in earlier studies (like [BN], where restriction to an infinite cylinder broke the symmetry thereby obscuring the criticality analysis. For a more recent analysis of this model see e.g. [LC]).

In our simulations there seemed to be clear phase segregation above the approximate threshold $p_c=0.98$. There a pair of sublattices take over i.e. the tiles increasingly concentrate on them deserting the two others. The pairing is forced by the structure of the densest packings above. 1-tile configurations on such a sublattice pair combined can together achieve arbitrarily good cover of ${\rm {\bf R}}^2$ and indeed tile it.

Figure 5b illustrates the near equilibrium configurations on the four sublattices in the supercritical regime ($p\approx 0.992$). The four possible sublattice matchings are the pairs next to each other in the same row or column in the $2\times 2$ array shown. Clearly here the two dominant ones are the ones on top row which superimposed to each other approximate fairly well the $p=1$ limit of a row shifted disordered tiling. 

\vskip .3truein
\noindent One should expect the voter mechanism to be more complex in the context of the ${\rm {\bf Z}}^2{\rm M}$ lattice. As we have seen, there any one of the four sublattices can support only a single densest packing (one of the maximally symmetric ones), whereas there are four pairs of sublattices that can each support an infinite number of densest packings. The criticality emerges from the system's decision on whether the horizontal or vertical slides are allowed in these packings.

\vskip .4truein
\centerline{\hbox{
 \psfig{figure=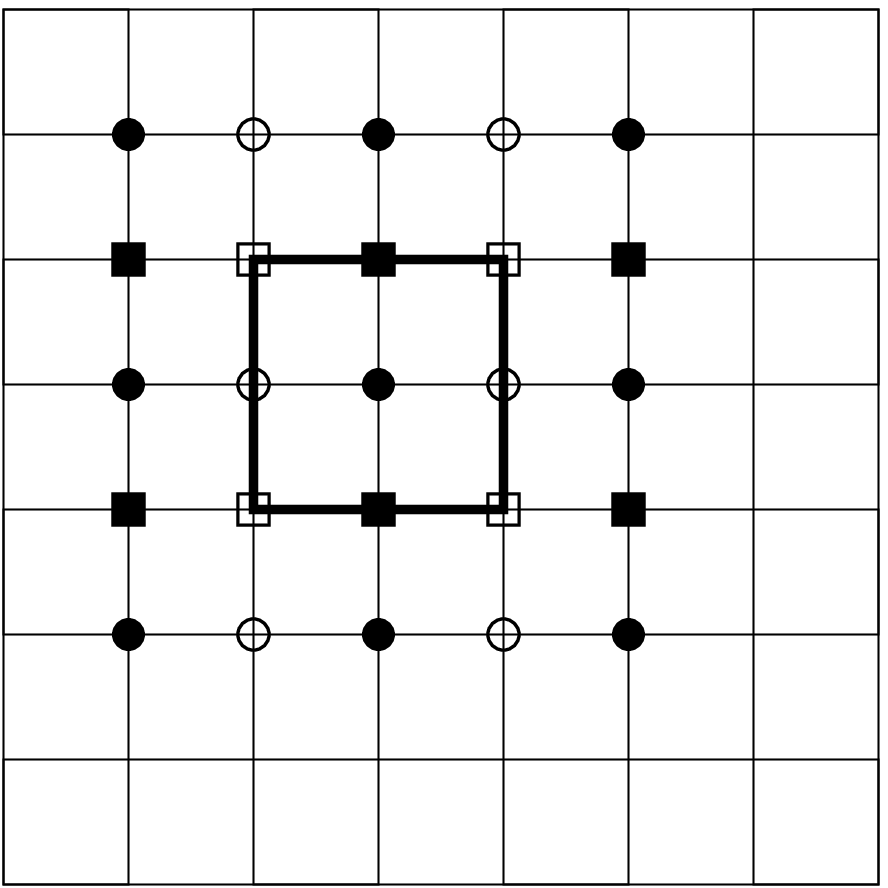,height=2.4in}
 \hskip .5truein
 \vbox{\psfig{figure=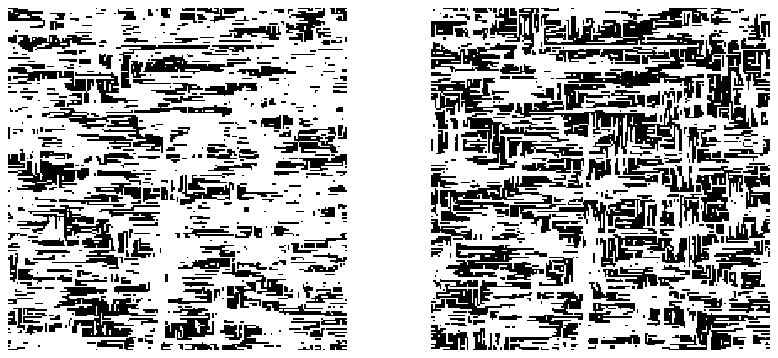,height=1.1in}
 \vskip .2truein
 \psfig{figure=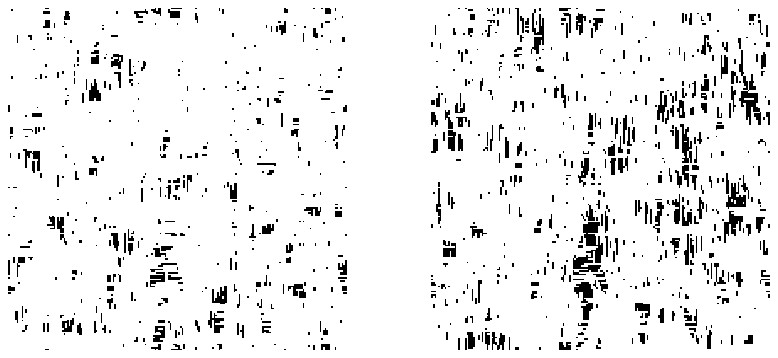,height=1.1in}
}
}}
\vskip .1truein
\centerline{{\bf Figure 5a, b.} ${\rm {\bf Z}^2}$ lattice with Moore neighborhood squared and a supercritical state.}
\centerline{($p=0.992$, $200\times 200$ torus, $10.000$ iterates from $B(1/8).$)}
\vskip .3truein

\noindent Figure 5a indicates the $25$-tuple from which the center site is updated under $f_p^2.$ Form now doublets from (horizontal) nearest neighbor dots (one black and one white). $f_1^2$ induces a unique update map on these, denote it by $f_{\bullet\circ}^2.$ The entries of the doublet $(\bullet\circ)$ as well as the update can be of the form $(0,0)$, $(0,1)$ or $(1,0).$ $f_{\bullet\circ}^2$ is again a map from $9$-tuples (of doubles) to $\{0,1\}^2$ and moreover by the Hard Core rule among the $6\times 3$ lattice points on which $f_{\bullet\circ}^2$ depends there are at most nine $1$'s.

The middle curve in Figure 4b gives the fraction of $3\times 3$ non-overlapping doublet neighborhoods with $k$ $1$'s in them that update to a doublet with $1$ in it. This map is identical to the three maps considered above, only now we map horizontal doublets instead of singletons (individual lattice site values). The reason that one must consider $f_{\bullet\circ}^2$ instead of $f_\bullet^2$ is that this map is independent of the horizontal slides in the configuration. It's graph indicates a weak majority voter character.

The rightmost curve in Figure 4b is what one gets by restricting $f_1^2$ to one sublattice alone. The map clearly fails to be majority voter-type. If one defines doublets by combining sublattices that are one horizontal and one vertical shift from each other the resulting map is not voter-like either. This is expected: this arrangement of $1$-tiles leaves gaps in the packing and cannot prevail under high packing pressure. 

\vskip .2truein
\noindent The above seems to support observed critical phenomena in the models. It shows that a majority voter mechanism operates at the level of the optimal sublattices. There are caveats though. As seen from the graphs the $f_1^2$ maps are not quite symmetric as they should be in the ideal voter rule. However this slight asymmetry actually favors the phase segregation. Spurious combinatorial phenomena may show up. There may e.g. be \lq\lq instability \rq\rq\ of the following kind: Consider $f_1^2$ on ${\rm {\bf Z}}^2$ and suppose the center entry in the $3\times 3$ diamond is $1$ and the rest of the entries are $0$'s. This neighborhood yields the update $1.$ But if the entry $1$ is anywhere else in the neighborhood while the others are still $0$'s the update will be $0.$ Whether details of this kind can influence the global relaxation of the PCA for $p<1$ will require further study.

\vskip .4truein
\noindent {\subtitle 2.2.4. Noncritical packing}
\vskip .3truein

\noindent The third and perhaps the most interesting class of packings on a lattice are the 2-d random packings. This class seems to coincide with the noncriticality of the PCA, a connection of definite interest. The analysis here concentrates on the Kagom\'e case but we believe that the same principles apply verbatim to the three cases among the Archimedean set and beyond. 

\vskip .2truein
\noindent Kagom\'e lattice ({\bf K}) is a $1/2$-thinning of the triangular lattice obtained by removing every other lattice line in all of the three lattice directions. It has appeared in the context of a number of statistical physics models ([B]) and it turns out that here, too, it yields some interesting insight in to the nature of the high density packings on lattices.

Each lattice point has four nearest neighbors in a \lq\lq hour glass\rq\rq\ formation. If we would choose the $1$-tile to be a hexagon with $N_x$ among its vertices these hexagons distributed according to the $1$'s from a legal configuration could never overlap, but they would not form a tiling either. However if one extends the hexagon to a rhombus with long axis aligned to the vertical axis of the hour glass, then these 1-tiles cannot ever overlap and moreover will form a tiling for the densest packing (a rhombus is two equilateral triangles glued together at their base). This construction immediately implies that the maximum density of $1$'s on the lattice is $1/3.$

The lattice divides in a natural way up to three identical sublattices, each of which is again a Kagom\'e lattice scaled by factor $\sqrt{3}$ and rotated appropriately as indicated in Figure 1 (the dots, circles and rings identify the three sublattices). Each of these sublattices supports a rhombus tiling. The arrangement on the left, three rhombi forming a hexagon, can be viewed as a double covering set since with it one can cover the plane with one rhombus overlap between neighboring hexagons. All the rhombi in this cover are supported by the same sublattice (the dots). These three covers correspond to the most symmetric of the densest packings on the Kagom\'e lattice.

The three-tuple of rhombi forming a hexagon can be rotated by $\pi.$ This {\bf flip} is an example of a local move. In the new formation the rhombi are centered on the ring sublattice. Hence in a domain containing $N$ non-overlapping hexagons there are at least $2^N$ different rhombus tilings. Indeed the exponential growth rate $h^{(2)}$ of rhombus tilings (entropy per tile) can be shown to be
$${1\over {4\pi^2}}\int_0^{2\pi}\int_0^{2\pi}\log{\left(1+e^{i\theta}+e^{i\phi}\right)}d\theta d\phi$$
(which cannot be evaluated in closed form but is approximately 0.32306, see [DMB], [W]). The set of rhombus tilings is closed under flips in the sense that any two tilings can be reached from each other through a finite sequence of flips (for a most general result in this direction see [P]).

The case at hand is very different from those in the previous sections including the ${\rm {\bf Z}}^2{\rm M}.$ In the last one the number of densest packings in a square of $n^2$ vertices is asymptotically $e^{h^{(1)}n}$ whereas in the Kagom\'e case it is $e^{h^{(2)}n^2}.$ The reason for this is that on ${\rm {\bf Z}}^2{\rm M}$ all the packings are still laminated whereas on Kagom\'e this is not the case: if weighted uniformly almost all of them are disordered without any periodic structures left in them. In other words in this class the densest packings are generically random packings.

The relaxation dynamics of the Hard Core PCA defined on Kagom\'e lattice is unlike in the cases above. There is no majority voter mechanism to favor any sublattice and there is no critical value for the packing pressure $p.$ In our simulation runs (about $200\times 200$ site tori) the relaxation appeared to be exponentially fast for all $p$ and the limiting configuration was evenly distributed on the three optimal sublattices. This is to be expected: the equilibrium measure should be the unique measure of maximal entropy since this is unique even at the $p=1$ limit (where it is the uniform measure on the optimal packings, additionally also invariant under the flips).

\vskip .2truein
\noindent Hard core on $(3.4.6.4)$ and $(3.12^2)$ are qualitatively similar to the above in high density packing. Figure 1 indicates densest packings with $\rho=1/3$. Both admit 2d random perturbations with the indicated local moves (arrows). We have computed the entropy bounds in Table I for $(3.4.6.4)$ from the independent choices on the period parallelogram (three inside the shaded indicated domain) and for $(3.12^2)$ from the rotations in density $1/2$ 12-gons in the period parallelogram.

\vskip .4truein
\noindent {\subtitle 2.2.5. Extensions: Union Jack and Quilt lattices}
\vskip .2truein

\noindent Union Jack (${\rm {\bf UJ}}$) is derived from the unit square lattice by adding the ascending and descending diagonals. It can also be obtained as a dual of the Archimedean lattice $(4.8^2)$ (i.e. it corresponds to the Laves tiling $[4.8^2],$ see [GS]). Quilt, ${\rm {\bf Q}}$, for which we do not know a more standard name, is the thinning of the Union Jack when one removes every other ascending and descending diagonal. These lattices are illustrated in Fig. 6a and b.

\vskip .3truein
\centerline{\hbox{
 \psfig{figure=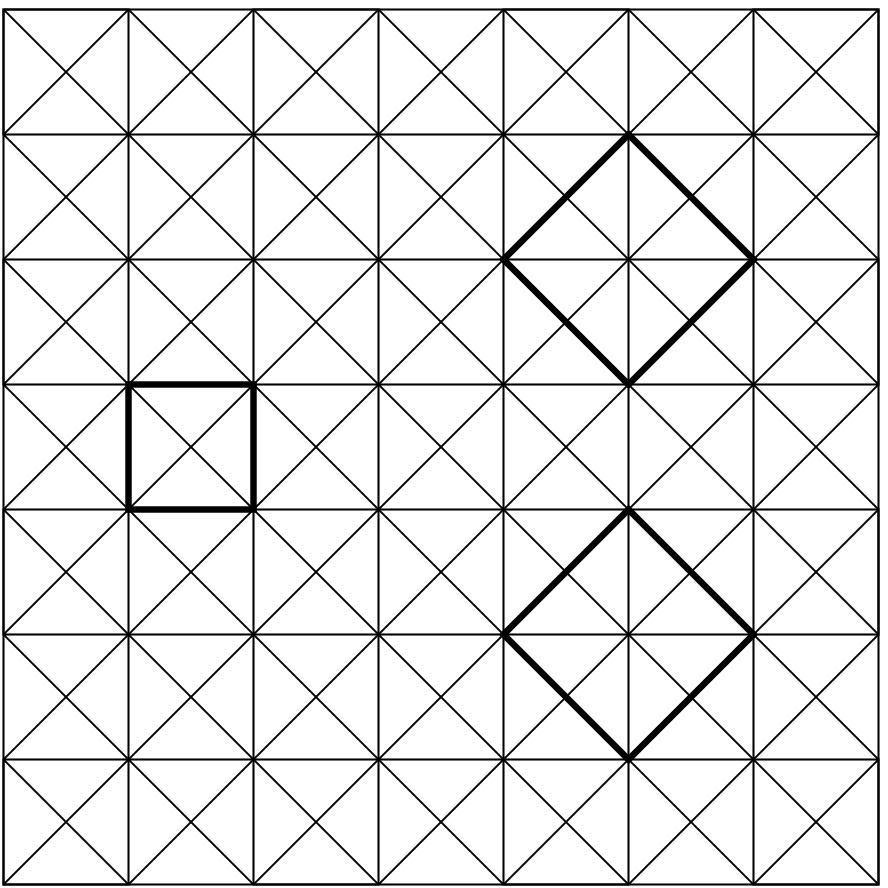,height=1.1in}
 \hskip .5truein
 \psfig{figure=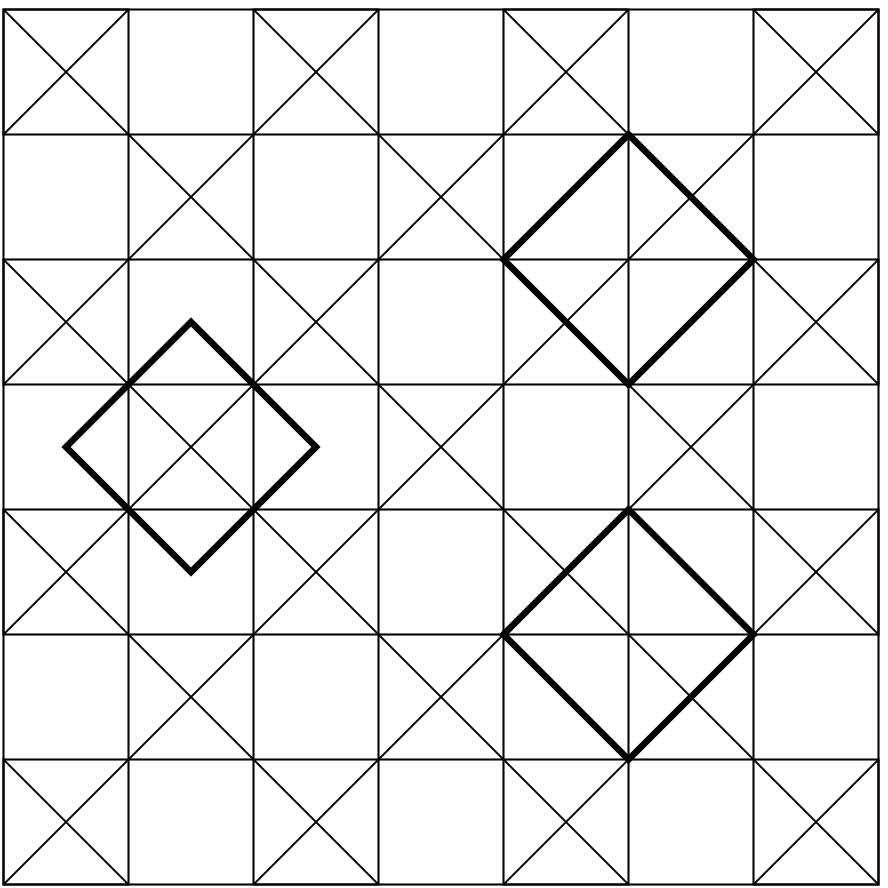,height=1.1in}
 \hskip .5truein
 \psfig{figure=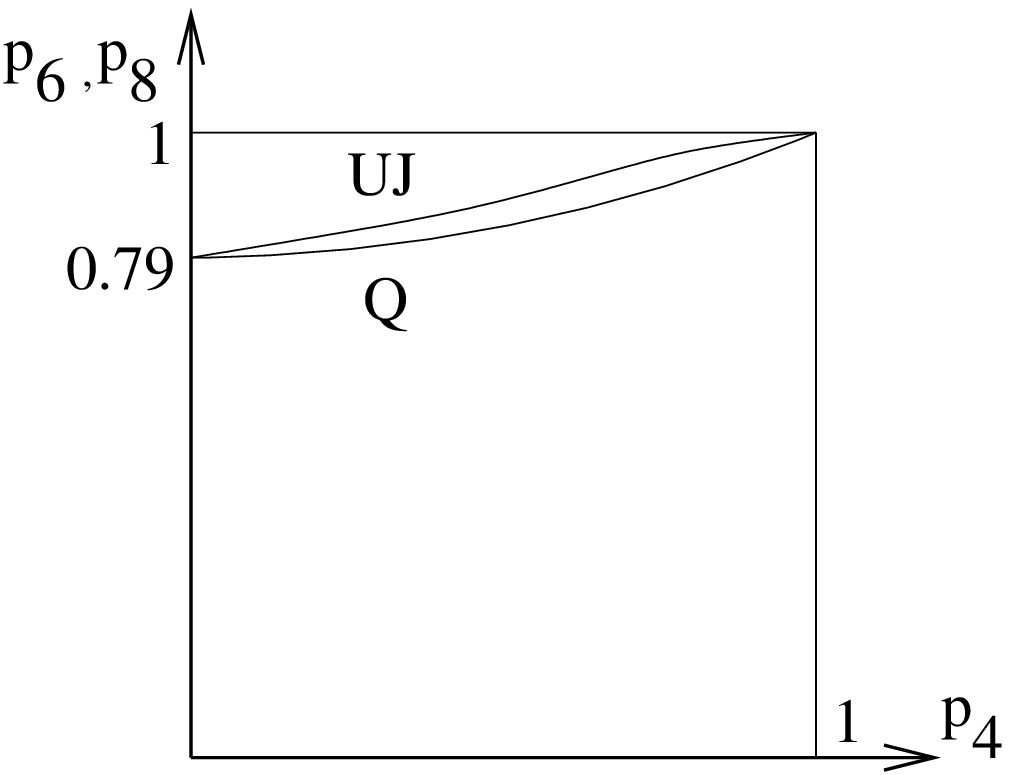,height=1.4in}
}}
\vskip .1truein
\centerline{{\bf Figure 6a, b, c.} Union Jack and Quilt graphs with tiles. Approximate critical curves.}
\vskip .2truein

\noindent The interest in these lattices stems from the fact that they refine ${\rm {\bf Z}}^2$ in two new ways and in the process bring along a novel type of 1-tile. In the case of the Union Jack lattice it is a unit square (hence having half the area of the diamonds) and in Quilt another diamond.

Our update rules will now have two parameters, the probabilities $p_4$ and either $p_8$ or $p_6$. The index refers to the degree of $x$, squares having four and diamonds eight/six neighboring sites on their boundaries in this lattice in Union Jack/Quilt respectively.

The model with $p_4=0$ is the Hard Square Model. The critical behavior on both lattice set-ups extends to the $p_4>0$ regime. In simulations the phase diagram appears approximately like in Figure 6c, the curves separating the sub- and supercritical phases. The curves should be above the $p_c\approx0.79$ horizontal line, increasing, reaching the northeast corner and be ordered as shown for the two models. Our simulations (on $100\times 100$ torus) support these claims, which can also be heuristically argued but we do not have a rigorous argument.

\vskip .2truein
\noindent From the packing point of view Union Jack lattice is somewhat ambiguous since the squares and diamonds do not have same area. In the high density limit $p_4, p_8\uparrow 1$ all-1-squares arrangement should prevail. Hence among the optimal packings there should be a unique densest one (off $p_8=1$). This behavior disappears in the Quilt lattice - the squares are there replaced by diamonds of degree four which are of equal area to the degree six diamonds. The optimal packings for the Quilt lattice are clearly the three sublattice tilings. No other perfect tilings exist in this set up. 

\vskip .2truein
\noindent The models on the Union Jack and Quilt lattices are for small $p_4$ just variants of the model on ${\rm {\bf Z}}^2.$ Consider again the $3\times 3$ diamond neighborhood on ${\rm {\bf Z}}^2.$ If in each of its $2\times 2$ subneighborhoods there is at least one symbol $1$ then on can compute that whereas $f_p^2$ yields update $1$ on ${\rm {\bf Z}}^2$ w.p. $p$, on Union Jack the update with $f^2_{\left(p_4,p_8\right)}$ will be $1$ at least w.p. $p_8(1-p_4)^4$ and on Quilt at least w.p. $p_6(1-p_4)^2.$ Hence the update probability on both of these lattices is near to the square lattice case.

\vskip .4truein
\noindent {\subtitle 3. Conclusions}
\vskip .2truein

\noindent In this paper we have studied the densest packings of 1's on certain planar lattices with the exclusion implemented by the nearest neighbor Hard Core Rule. The choice of uniform lattices was paramount: we wanted to investigate the most natural discrete analog of the densest packing of disks/spheres/hyperballs in the Euclidean world.

As we have seen a rather striking picture emerges; either there are only a few densest packings or there is an exponential number of them (in the size of the domain). This split is a consequence of the densest packing being either rigid or loose (i.e. the ground state having zero or positive 1d or 2d residual entropy as for example in the Ice model of Statistical Mechanics). This sparseness or richness of the set of the densest configurations in turn is a direct consequence of the existence of a local move. The lack of a local move implies long range order whereas the existence of a local move enables one to join densest finite patches into a densest global configuration.

The packing dynamics can be critical and the packing types are intimately related to this criticality/noncriticality in the generating PCA. In all our cases the lattices with rigid ground states (no local move) are approached via a critical transition when the packing pressure increases. Conversely when the set of ground states is rich enough, they can be piecewise glued together to a global one and a critical transition is not imposed on the model in the quenching of the PCA.

The processes involved here are quite subtle which is perhaps best indicated by the existence of the borderline critical case of slide packings. These are likely not to be combinatorial oddities - indeed they resemble a great deal the well known Barlow packing, the optimal packing of 3-d hard spheres (see [CG-SS]). In this context there is no local move but there is an infinite slide move. Hence there is limited 1-d long range order and apparently a phase transition in the packing pressure that goes with it. There has been attempts to clarify this using e.g. the Pirogov-Sinai theory to no avail (private communication). To fully understand the nature of the transitions in these models more studies are needed.

\vskip .4truein
\noindent {\subtitle Acknowledgements}
\vskip .2truein

\noindent The author would like to thank Joel Lebowitz for discussions on these models and the referees for criticism that led to improvement and in particular to the completion of Table I. 

\vfill
\eject

\vskip .4truein
\noindent {\subtitle References}

\vskip .2truein

\item{[B]} Baxter, R.J.:{\sl Exactly solvable models in statistical mechanics},
Academic Press, 1982.

\item{[BET]} Baxter, R.J., Enting, I. G., Tsang, S. K., {\sl J. of Stat. Phys.}, {\bf 22},
pp. 465-89, 1980.


\item{[BN]} Bellemans A., Nigam R. K.: Phase transitions in the hard-square lattice gas, {\sl Phys. Rev. Lett.} {\bf 16}, pp. 1038-9, 1966.


\item{[BW]} Brightwell, G., Winkler, P.: A second threshold for the Hard Core Model in a Bethe lattice, {\sl Random Structures and Algorithms} {\bf 24} (3), pp. 813-14, 2004.

\item{[CE]} Cohn, H., Elkies, N.: New upper bounds on sphere packings I, {\sl Ann. Math.} {\bf 157}, pp. 689-714, 2003.

\item{[CG-SS]} Conway, J.H., Goodman-Strauss, C., Sloane, N.J.A.: Recent progress in sphere packing, {\sl Current Developments in Mathematics}, editors B. Mazur, W. Schmid, S. T. Yau, D. Jerison, I. Singer and D. Stroock, Cambridge, pp. 37-76, 1999. 

\item{[DMB]} Destainville, N., Mosseri, R., Bailly F: Configurational entropy of codimension-one tilings and directed membranes, {\sl J. of Stat. Phys.}, {\bf 87}, pp. 697-754, 1997.
 
\item{[DS]} Dobrushin, R., Shlosman, S., in {\sl Statistical Physics and Dynamical Systems},
J. Fritz, A. Jaffe and D. Szasz, eds., Birkh\"auser, pp. 347-370, 1985.

\item{[E]} Eloranta, K.: A note on certain rigid subshifts, {\sl
Ergodic Theory of ${\rm {\bf Z}}^d$-Actions}, London Mat. Soc. Lect. Notes {\bf 228}, Cambridge Univ. Press, pp. 307-317, 1996.



\item{[GK]} Galvin, D., Kahn, J.: On phase transition in the Hard-Core Model on ${\rm {\bf Z}}^d$, {\sl Combinatorics, Probability \& Computing} {\bf 13} (2), pp. 137-64, 2004.

\item{[GS]} Gr\"unbaum, B., Shephard, G. C.: {\sl Tilings and patterns}, Freeman, 1987.

\item{[L]} Liggett, T.: {\sl Interacting Particle Systems}, Springer, 1985.

\item{[LC]} Lafuente, L., Cuesta, J. A.: Phase behavior of hard-core lattice gases: A fundamental measure approach, {\sl J. Chem. Phys.} Vol {\bf 119}, pp. 10832-43, 2003. 

\item{[MSS]} Milosevic, S., Stosic, B., Stosic, T.: Towards finding
exact residual entropies of the Ising ferromagnets, {\sl Physica A},
{\bf 157}, pp. 899-906, 1989.

\item{[P]} Propp, J.: Lattice structure for orientations of graphs, {\tt arxiv.org/abs/math.CO/0209005}.

\item{[R]} Runnels, L. K.: Phase transitions of hard sphere lattice gasses, {\sl Comm. Math. Phys.} {\bf 40}, pp. 37-48, 1975.

\item{[RS]} Radulescu, D., Styer, D.: The Dobrushin-Schlosman Phase Uniqueness Criterion and Applications to Hard Squares, {\sl J. of Stat. Phys.}, {\bf 49}, 281-95, 1987.

\item{[W]} Wannier, G. H.: Antiferromagnetism. The triangular Ising net, {\sl Phys. Rev.} {\bf 79}, pp. 357-64, 1950 and {\sl Phys. Rev. B } {\bf 7}, 5017, 1973.
\vfill
\eject


\voffset=1truein
\hoffset=-.35truein

{
\offinterlineskip
\tabskip=0pt
\halign{ 
\vrule height2.75ex depth1.25ex width 0.6pt #\tabskip=1em &
#\hfil &\vrule \hfil # \hfil &  #\hfil &\vrule # &\hfil # \hfil &\vrule # & \hfil #\hfil &\vrule # & #\hfil & #\vrule width 0.6pt \tabskip=0pt\cr
\noalign{\hrule height 1pt}
&  &&  &&  &&  &&  &\cr
& Graph && Optimal subgraphs, && Pack.  && $\rho$ && Critical pressures \& densities, &\cr
&  && multiplicity, tiles && type  &&  && residual entropies &\cr
&  &&  &&  &&  &&  &\cr
\noalign{\hrule height 1pt}

& $(4^4)$ (${\rm {\bf Z}}^2$) && $\sqrt{2}\ {\rm {\bf Z}}^2$, 2,  diamond && L && $1/2$ && $p_c\approx 0.79,\ \rho(p_c)\approx 0.36$ &\cr
\noalign{\hrule}

& $(6^3)$ (${\rm {\bf H}}$) && $\sqrt{3}\ {\rm {\bf T}}$, 2, hexagon && L && $1/2$ && $p_c\approx 0.87,\ \rho(p_c)\approx 0.4$  &\cr
\noalign{\hrule}

& $(3^6)$ (${\rm {\bf T}}$) && $\sqrt{3}\ {\rm {\bf T}}$, 3, hexagon && L && $1/3$ && $p_c\approx 0.90,\ \rho(p_c)\approx 0.26$ &\cr
\noalign{\hrule height 1pt}

& $(4.8^2)$ && $\simeq (3^2.4.3.4)$, 2, 5-gon && L && $1/2$ && $p_c\approx 0.90,\ \rho(p_c)\approx 0.4$ &\cr
\noalign{\hrule}

& $(4.6.12)$ && $\simeq (3^4.6)$, 2, 5-gon && L && $1/2$ && $p_c\approx 0.91,\ \rho(p_c)\approx 0.42$ &\cr
\noalign{\hrule}

& $(3^2.4.3.4)$ && $\simeq {\rm {\bf T}}$, 3, 6-gon && L && $1/3$ && $p_c\approx 0.99$, $\rho(p_c)\approx 0.3$ &\cr
\noalign{\hrule}

& $(3^4.6)$ && $\sqrt{3}\ (3^4.6)$, 3, 5-gon && L && $1/3$ && $p_c\approx 0.97,\ \rho(p_c)\approx 0.29$ &\cr
\noalign{\hrule}

& $(3^3.4^2)$ && $3\ {\rm {\bf Z}}$ stack, $\infty$ && RL && $1/3$ && $h^{(1)}={1\over 2}\log{2}$ &\cr
\noalign{\hrule}

& $(3.4.6.4)$ && $\simeq (3^4. 6),\ \infty$, 5-gon && R && $1/3$ && $h^{(2)}\ge {3\over {16}}\log{2}$ &\cr
\noalign{\hrule}

& $(3.6.3.6)$ (${\rm {\bf K}}$)&& $\sqrt{3}\ (3.6.3.6),\ \infty$, rhombus && R && $1/3$ && $h^{(2)}\approx 0.323$ &\cr
\noalign{\hrule}

& $(3.12^2)$ && $\simeq (3^4.6),\ \infty$, 5-gon && R && $1/3$ && $h^{(2)}\ge {1\over {18}}\log{2}$  &\cr
\noalign{\hrule height 1pt}

& ${\rm {\bf Z}}^2{\rm M}$ && $2\ {\rm {\bf Z}}$ stack, $\infty$, square && RL && $1/4$ && $p_c\approx 0.98?,\ h^{(1)}={1\over 2}\log{2}$&\cr
\noalign{\hrule}

& $[4.8^2]$ ({\bf UJ}) && ${\rm {\bf Z}}^2$, 1, sq., $\sqrt{2}\ {\rm {\bf Z}}^2$, 2, dia. && L && $1/2$ && increasing critical curve &\cr
\noalign{\hrule}

& Quilt ({\bf Q}) && $\sqrt{2}\ {\rm {\bf Z}}^2$, 3, diamond && L && $1/3$ && increasing critical curve &\cr
\noalign{\hrule height 1pt}
}}

\vskip .6truein
\noindent \item{} {\bf Table I}. Packing properties of the Archimedean and other graphs. The first three lines correspond to the regular graphs, the first 11 to the Archimedean ones and last three are non-Archimedean.

\vskip .1truein
\noindent \item{$\bullet$} Optimal subgraphs: the named ones carry the most symmetric densest packings. $\simeq$ means topological equivalence.
\item{$\bullet$} Multiplicity: number of densest packings. Finite number means that all optimal ones are isometric to the most symmetric one, infinite that there are random perturbations.
\item{$\bullet$} Tile: $n$-gon is not regular, whereas diamond, square and hexagon are. Rhombus forms from two equilateral triangles glued together at base. For further details see Section 2.1. 
\item{$\bullet$} Type: L = laminated packing, RL = random laminated packing i.e. 1d slide packing, R = 2d random packing.
\item{$\bullet$} $\rho$: density of the optimal packings in the full graph.
\item{$\bullet$} $h^{(1)},\ h^{(2)}$: the one- and two-dimensional residual entropies of the densest packings.  

\vfill
\eject

\end